\pdfoutput=1
\documentclass[11pt,twoside,a4paper,cmspaper,final,collab]{cms-tdr}
\def\svnVersion{c27ae8d}\def\svnDate{2024/02/23}\def\cmsCernNoTag{CERN-EP-2023-205}\def\cmsCernDate{\today}\def\cmsMessage{Published in the Journal of Instrumentation as \href{http://dx.doi.org/10.1088/1748-0221/19/02/P02031}{\doi{10.1088/1748-0221/19/02/P02031}.}}
\begin{document}\cmsNoteHeader{MUO-22-001}

\newenvironment{cmslist}{\begin{list}{$-$}{%
    \setlength{\topsep}{0.0ex}%
    \setlength{\partopsep}{0.0ex}%
    \setlength{\parsep}{0.5\parskip}%
    \setlength{\itemsep}{0.0ex}%
    \setlength{\leftmargin}{3ex}%
    \setlength{\rightmargin}{0ex}%
}}{\end{list}}

\newcommand{\ttH}{\ensuremath{\ttbar\PH}\xspace}
\newcommand{\pp}{\ensuremath{\Pp\Pp}\xspace}
\newcommand{\Zjets}{\ensuremath{\PZ+\text{jets}}\xspace}
\newcommand{\Wjets}{\ensuremath{\PW+\text{jets}}\xspace}

\newcommand{\DeepJet}{\ensuremath{\textsc{DeepJet}}\xspace}
\newcommand{\Run}[1]{\mbox{Run #1}\xspace}

\newcommand{\Npass}{\ensuremath{N_{\text{pass}}}\xspace}
\newcommand{\Nfail}{\ensuremath{N_{\text{fail}}}\xspace}
\newcommand{\abseta}{\ensuremath{\abs{\eta}}\xspace}
\newcommand{\chisq}{\ensuremath{\chi^2}\xspace}
\newcommand{\ptmu}{\ensuremath{\pt^{\PGm}}\xspace}
\newcommand{\pmu}{\ensuremath{p^{\PGm}}\xspace}
\newcommand{\Irel}{\ensuremath{I_{\text{rel}}}\xspace}
\newcommand{\ptrel}{\ensuremath{\pt^{\text{rel}}}\xspace}
\newcommand{\mtfix}{\ensuremath{\mT^{\text{fix}}}\xspace}
\newcommand{\ptfix}{\ensuremath{\pt^{\text{fix}}}\xspace}

\cmsNoteHeader{MUO-22-001}
\title{Muon identification using multivariate techniques in the CMS experiment in proton-proton collisions at \texorpdfstring{$\sqrt{s}=13\TeV$}{sqrt(s) = 13 TeV}}

\date{\today}

\abstract{The identification of prompt and isolated muons, as well as muons from heavy-flavour hadron decays, is an important task. We developed two multivariate techniques to provide highly efficient identification for muons with transverse momentum greater than 10\GeV. One provides a continuous variable as an alternative to a cut-based identification selection and offers a better discrimination power against misidentified muons. The other one selects prompt and isolated muons by using isolation requirements to reduce the contamination from nonprompt muons arising in heavy-flavour hadron decays. Both algorithms are developed using 59.7\fbinv of proton-proton collisions data at a centre-of-mass energy of $\sqrt{s}=13\TeV$ collected in 2018 with the CMS experiment at the CERN LHC.}

\hypersetup{
pdfauthor={CMS Collaboration},
pdftitle={Identification of prompt and isolated muons using multivariate techniques at the CMS experiment in proton-proton collisions at sqrt(s) = 13 TeV},
pdfsubject={CMS},
pdfkeywords={CMS, muon}}

\maketitle

\section{Introduction}

Accurate measurements of processes with muons in the final state are among the main goals of the CMS experiment~\cite{Chatrchyan:2008zzk} at the CERN LHC. Because many analyses rely on the presence of these leptons to identify interesting physical processes and suppress backgrounds, a successful physics program relies on a high-quality muon selection. First, it is necessary to discriminate genuine muons from other background sources that fake muons, such as spurious hits misreconstructed as muons, or misidentified charged hadrons. Then, for analyses targeting prompt muons from decays of the \PW, \PZ, or Higgs bosons, or muons from \PGt lepton decays, it is important to avoid a contamination from nonprompt muons produced by hadron decays, especially \PQb quark hadrons.

The LHC \Run2 at CERN occurred during the years 2015 to 2018 and provided an unprecedented number of proton-proton (\pp) collision events at a centre-of-mass energy, $\sqrt{s}$, of 13\TeV recorded by the CMS experiment. During this period, muon identification relied on a cut-based approach, using a set of requirements on individual variables related to the information from the tracker and the muon system. This approach defines three sets of selection criteria~\cite{CMS:2018rym} with a differing balances between efficiency and purity that allow each analysis to choose the desired set. However, whenever the purity of the selected sample is critical,
or the muon identification performance
is degraded due to the number of simultaneous collisions in the same or adjacent bunch crossings (pileup, PU), the use of more sophisticated techniques is crucial to maintain an acceptable compromise between efficiency and purity.

We adopted a multivariate analysis (MVA) approach to design two muon identification discriminators. The first discriminator, referred to as ``muon MVA ID'', is used for generic muon identification while minimizing hadron misidentification. This discriminator has been trained using muon identification variables defined in Ref.~\cite{CMS:2018rym} and is presented in this paper for the first time. The second MVA, referred to as the ``prompt-muon MVA'', is designed to accurately identify isolated prompt muons, for those particular cases in which it is critical to tag such muons, particularly those produced from the decays of \PW, \PZ, Higgs bosons, or \PGt leptons.

The prompt-muon MVA discriminator was trained in the context of multileptonic analyses with high yields of nonprompt backgrounds, and it was already used in several CMS analyses with multilepton final states~\cite{CMS:2020mpn,CMS:2018fdh,CMS:2018uxb,CMS:2021pqj,CMS:2021icx,CMS:2019efc,CMS:2021cox,CMS:2023tttt}. This prompt-muon MVA was designed for isolated muons and it was trained using both muon identification and isolation variables. Hence, it cannot be used to identify other genuine muons, such as muons from heavy-flavour decays.

Following a similar MVA approach, a second classifier was trained, combining only muon identification variables, to construct a more general and robust discriminator able to outperform the cut-based identification~\cite{CMS:2018rym}. Such a discriminator could offer a continuous variable that provides more flexibility to pick the desired trade-off between signal and background efficiencies. With these goals in mind, we developed the muon MVA ID discriminator for muons with transverse momentum (\pt) $>$10\GeV and plan to use it for the \Run3 era of the LHC running. Low-\pt muons come from low-mass resonances, so they present different signatures when compared to muons from \PZ bosons, hence they require special treatment and are beyond the scope of this study.  For identification of low-\pt muons, a special MVA-based discriminator, referred to as ``Soft MVA ID''~\cite{CMS:2019bbr}, was developed and used during \Run2. It was trained using the 2016 data set for muons with $\pt < 10\GeV$. We present technical aspects about the training of the two algorithms in this paper and compare their performance in data and simulated events.

The paper is organized as follows: Section~\ref{sec:cms} briefly describes the CMS experiment, and Section~\ref{sec:samples} summarizes the data and simulated samples used in this work. Section~\ref{sec:muons} contains the details of the event reconstruction. Section~\ref{sec:mva} provides a
full description of the MVA discriminators, together with the details of their training. Section~\ref{sec:run2perf}
discusses the performance of the algorithms on the \Run2 data and Monte Carlo (MC) simulation. The results are summarized in Section~\ref{sec:summary}.

\section{The CMS detector}
\label{sec:cms}

The central feature of the CMS apparatus is a superconducting solenoid of 6\unit{m} internal diameter, providing a magnetic
field of 3.8\unit{T}. Within the solenoid volume there are a silicon pixel and strip tracker, a lead tungstate crystal
electromagnetic calorimeter, and a brass and scintillator hadron calorimeter, each composed of a barrel
and two endcap sections. Forward calorimeters extend the pseudorapidity ($\eta$) coverage provided by the barrel and endcap detectors.
Muons are measured in gas-ionization detectors embedded in the steel flux-return yoke outside the solenoid. A more detailed
description of the CMS detector, together with a definition of the coordinate system used and the relevant kinematic
variables, is reported in Ref.~\cite{Chatrchyan:2008zzk}.

\begin{figure}[!ht]
  \centering
  \includegraphics[width=0.8\textwidth]{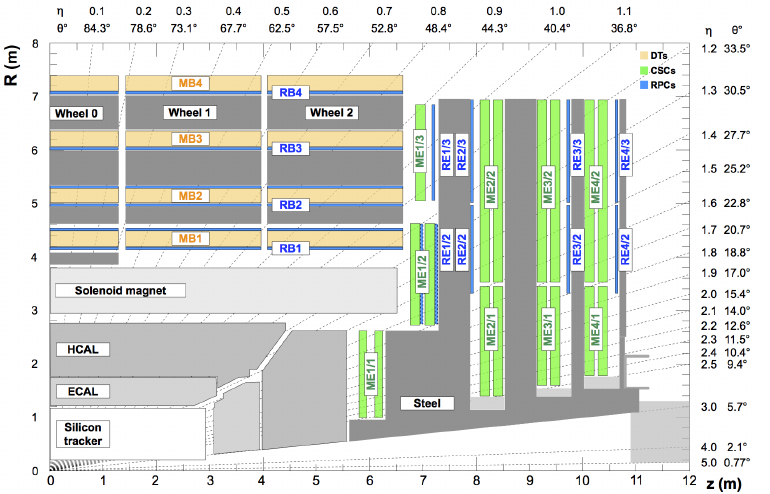}
  \caption{Layout of a quadrant of the CMS detector with the muon system highlighted~\cite{CMS:2018rym}. The chambers are named
MB for muon barrel, corresponding to DT chambers, and ME for muon endcap,
corresponding to CSC chambers. The RPC chambers are named RB and RE for barrel
and endcaps, respectively.}
  \label{fig:cms}
\end{figure}

The CMS muon system consists of three types of gas-ionization detectors: drift tube chambers (DTs), cathode strip chambers (CSCs),
and resistive-plate chambers (RPCs).
The DT and CSC detectors are located in the regions of $\abseta < 1.2$
and $0.9 < \abseta < 2.4$, respectively, and are complemented by the RPCs in the range $\abseta < 1.9$. The chambers are
arranged to maximize the coverage and to provide some overlap wherever possible. In both the barrel and endcap regions,
the chambers are grouped into four ``muon stations'', separated by the steel absorber of the flux-return yoke. A detailed description of these detectors,
including the gas composition and operating voltage, is reported in Ref.~\cite{CMS:2018rym}. Figure~\ref{fig:cms} shows a schematic representation of the CMS detector.

Events of interest are selected using a two-tiered trigger system. The first level, composed of custom hardware processors,
uses information from the calorimeters and muon detectors to select events at a rate of around 100\unit{kHz} within a fixed
latency of about 4\mus~\cite{CMS:2020cmk}. The second level, known as the high-level trigger, consists of a farm of
processors running a version of the full event reconstruction software optimized for fast processing, and reduces the event
rate to around 1\unit{kHz} before data storage~\cite{CMS:2016ngn}.

\section{Data and simulated samples}
\label{sec:samples}

The studies described in this paper are based on a data set of \pp collisions at $\sqrt{s} = 13\TeV$ produced at the LHC in 2018 and corresponding to an integrated luminosity of 59.7\fbinv~\cite{CMS-PAS-LUM-18-002, CMS:2021lum} recorded using the CMS experiment. The 2018 data set is representative of the \Run2 data-taking conditions. The trigger required the presence of a muon with a $\pt > 24\GeV$ threshold fulfilling loose isolation requirements~\cite{CMS:2021yvr}.

Events are simulated representing the main standard model (SM) processes and matching the detector and PU conditions of the 2018 data-taking period. Drell--Yan (DY) dilepton events are generated at next-to-leading order (NLO) in perturbative quantum chromodynamics (QCD) using the \MGvATNLO 2.6.1 generator~\cite{Alwall:2014hca, Alwall:2007fs}.
The \POWHEG 2.0~\cite{Alioli:2010xd,Frixione:2007vw,Nason:2004rx} generator at NLO accuracy in QCD is used to simulate top quark pair production (\ttbar) and the production of a Higgs boson in association with a top quark pair (\ttH). Samples of \Zjets, \Wjets, and multijet events are generated at leading order (LO) in QCD using the \MGvATNLO generator. For all the MC simulated samples, the primary interaction is overlaid with additional simulated minimum-bias events to simulate the effect of the PU. The  NNPDF 3.1 next-to-NLO~\cite{Ball:2017pdf} parton distribution function set is used for generating all MC samples. Parton showering and hadronization are simulated with \PYTHIA~v8.240~\cite{Sjostrand:2014zea,Skands:2014pea}
using the underlying event tune CP5~\cite{Sirunyan:2019dfx}. In the event simulation
using \MGvATNLO, the FxFx~\cite{Frederix:2012ps} (MLM ~\cite{Alwall:2007fs}) merging scheme is used for NLO
(LO) samples, to avoid possible double-counting of jets from the matrix element calculations and parton shower. The generated events are then processed through a detailed simulation of the CMS detector based on \GEANTfour 10.04~\cite{geant4} and are reconstructed with the same algorithms as used for data.

For the training and misidentification rate studies for the muon MVA ID, two different \ttbar samples that include semileptonic decays are used. The simulated sample used to check the performance includes generation-level information for the particles produced in the PU interactions, whereas the other sample does not. The choice of a \ttbar sample for the training provides a wide variety of sources of genuine muons, including muons from prompt decays and from heavy-flavour hadron decays.  To determine the source of muons in simulated events we look for the geometrical matching between the hits of the simulated  muon track and those of the muon track reconstructed in the muon system. From the simulated muon track we are able to trace the generated parent particle. In the simulated \ttbar sample used to check the performance, we find that 60\% of muons with $\pt >10\GeV$ passing the loose cut-based identification criteria~\cite{CMS:2018rym} are prompt isolated muons originating from the primary interaction (gauge boson decays); 8\% are isolated muons from \PGt lepton decays; 30\% are nonisolated muons from heavy-flavour decays (\PQb and \PQc hadrons); and 2\% are muons from light-hadron decays (pions or kaons) or nonprompt muon candidates from hadron misreconstruction.

\section{Muon reconstruction, identification, and isolation}
\label{sec:muons}

In the standard CMS reconstruction procedure for \pp collisions, muon tracks are first reconstructed independently in the inner tracker and in
the muon systems. In the muon system, tracks called ``standalone muons'' are reconstructed by using information from the
DT, CSC, and RPC detectors along the muon trajectory using the Kalman filtering technique~\cite{Fruhwirth:1987fm}. Within
each station, multiple detector planes record muon hits. The hits within a station are combined into segments, which
are in turn fitted into standalone-muon tracks. If the momentum, direction, and position in the transverse plane of a standalone-muon tracks and a track reconstructed in the inner tracker~\cite{CMS:2014pgm} are compatible, then a global track is fitted by combining hits from the tracker track and standalone-muon track in a common fit using the Kalman filtering method, referred to as ``global muons''.
A third type of muons referred to as ``tracker muons''
are reconstructed by extrapolating the inner tracker tracks outward to the muon system with loose geometrical matching to the DT or
CSC segments, these muons need to have a \pt of at least 3\GeV, otherwise they would not be able to reach the first station in the muon system. If at least one muon segment matches the extrapolated track, the track is qualified as a ``tracker muon''. Tracker muons have higher efficiency than global muons for low muon momenta and in regions of the CMS detector that are less instrumented.
More details about muon reconstruction and the corresponding track parameter resolution of each muon type are reported in Refs.~\cite{CMS:2018rym, CMS:2019ied}.

The primary vertex (PV) is taken to be the vertex corresponding to the hardest scattering in the event, evaluated using tracking
information alone, as described in Section 9.4.1 of Ref.~\cite{Contardo:2020886}. The performance of the PV determination depends on the \pt of the particles, as described  in Ref.~\cite{CMS:2014pgm}.

Reconstructed muons are an input into the CMS particle-flow (PF) algorithm~\cite{CMS-PRF-14-001}, which reconstructs and identifies
each individual particle in an event by combining the information from all the CMS subdetectors.
Particles reconstructed by the PF algorithm are classified into charged and neutral hadrons, photons, electrons, and muons.

After reconstruction, hadrons are clustered into jets using the anti-\kt algorithm~\cite{Cacciari:2008gp, Cacciari:2011ma}
with a distance parameter of 0.4. For this analysis, only jets with $\pt>10\GeV$ are selected. Charged hadrons not associated with the PV are excluded from the clustering. The energy of the reconstructed jets is corrected for residual PU effects using the method described in Refs.~\cite{Cacciari:2008gn, Cacciari:2007fd} and calibrated as a function of jet \pt and $\eta$~\cite{Khachatryan:2016kdb}. Jets originating from the hadronization of \PQb quarks are identified with the \DeepJet \PQb tagging algorithm~\cite{CMS-DP-2023-005,Bols:2020bkb}. We define the nearest jet to the muon as the one in which the muon candidate is clustered.

In the following we describe the standard CMS muon identification method, which makes use of a set of variables related to the muon track candidate information from both the tracker and the muon system. The most common sets of selection criteria are referred to as loose, medium, and tight. Their efficiencies, as measured during the \Run2 data taking by selecting muons with $\pt>20\GeV$, are 99.75, 98.25, and 96.00\% with statistical uncertainties between 0.02 and 0.03\% for muons with $\abseta < 0.9$. Then, for muons with $\abseta > 0.9$ the efficiencies for each WP are 99.77, 98.55, and 97.46\% with the statistical uncertainties between 0.02 and 0.04\% ~\cite{CMS:2018rym}.

The ``relative isolation'' variable is used to distinguish between muons from heavy-flavour decays and the prompt muons. It is defined as the \pt sum of all charged hadrons, the transverse energies of all photons and of all neutral hadrons reconstructed by the PF algorithm in a cone of angle $\DR = \sqrt{\smash[b]{(\Delta\phi)^2+(\Delta\eta)^2}}$ around the muon direction, divided by the muon \pt, where $\phi$ is the azimuthal angle.
The charged PF particles not associated with the PV are not included in this sum, whereas the contribution from neutral particles arising from PU is taken into account by using the following two alternative approaches~\cite{CMS:2018rym}. In the first approach, the corrected energy sum from neutral particles is obtained by subtracting the sum of charged-hadron deposits originating from PU vertices scaled by a factor of 0.5 from the energy sum of neutral hadrons and photons~\cite{CMS:2018rym}. In the second approach, the sums are corrected using the average energy density in the event ($\rho$)~\cite{CMS:2015xaf}, scaled by the corresponding ``effective area'' ($\mathcal{A}$). If the correction exceeds the PF cluster sum, the correction to the isolation is set to zero. Effective areas are determined independently for the electromagnetic and the hadronic energy sums and separately in barrel and endcap components.

For this study, two isolation variables are used: one computed with all the PF particles within an $\DR = 0.4$ cone and corrected using the first method, and the other, referred to as ``mini-isolation''~\cite{CMS:2016krz_miniiso}, for which the cone size varies as a function of the muon \pt: $\DR = 0.2$ when $\pt<50\GeV$, $\DR = 10\GeV/\pt$ when $50<\pt<200\GeV$, and $\DR = 0.05$ otherwise, corrected using the second method. This variable is particularly well suited to identify isolated muons in events with significant hadronic activity or in the Lorentz-boosted regime where accidental overlaps of muons and jets may occur, but still performs just as well in cleaner environments.

\section{Definition of the multivariate discriminators}
\label{sec:mva}

Two classifiers using MVA techniques are presented. Muons used in the training of both classifiers must pass the loose cut-based identification criteria and have $\pt > 10$ (5)\GeV for the muon MVA ID (prompt-muon MVA). The loose identification criteria requires muons selected by the PF algorithm to be either global muons or tracker muons. This preselection shows an efficiency of almost 100\% and helps to avoid poorly reconstructed muons that are not relevant for analyses.

\subsection{General muon identification}
\label{sec:mvaid}

The muon MVA ID discriminant is trained to distinguish genuine (\ie signal) muons from background muons. Signal muons are produced promptly in the decay of \PW, \PZ, and Higgs bosons, isolated muons from \PGt leptons decays, and nonisolated muons from heavy-flavour hadron decays. Background muons come from light hadron (\ie kaons and pions) decays and spurious signatures in the detector misreconstructed as muons,
such as hadrons detected in the muon system after traversing the calorimeters and the steel flux-return yoke.

\begin{figure}[!ht]
  \centering
  \includegraphics[width=0.42\textwidth]{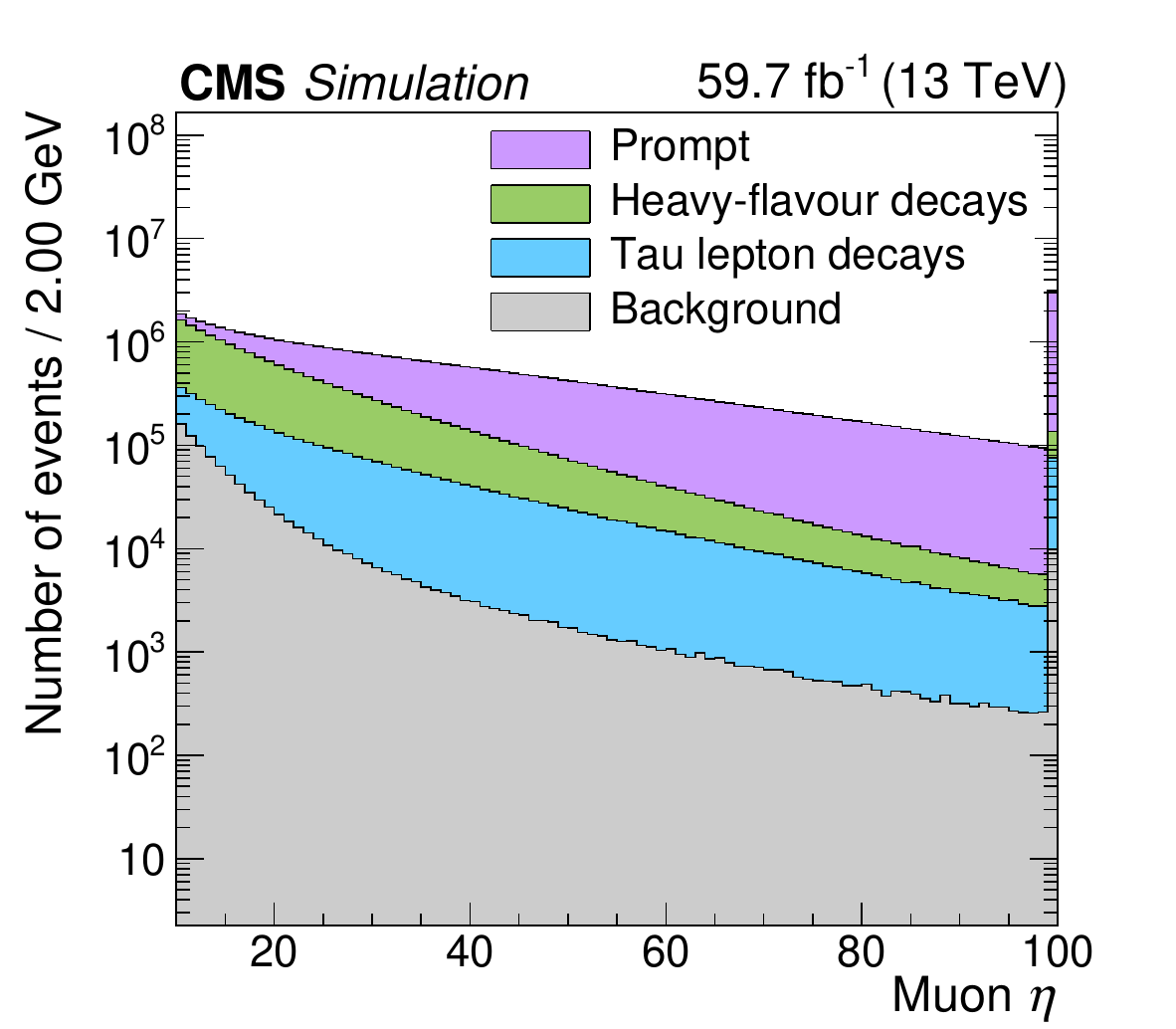}
  \includegraphics[width=0.42\textwidth]{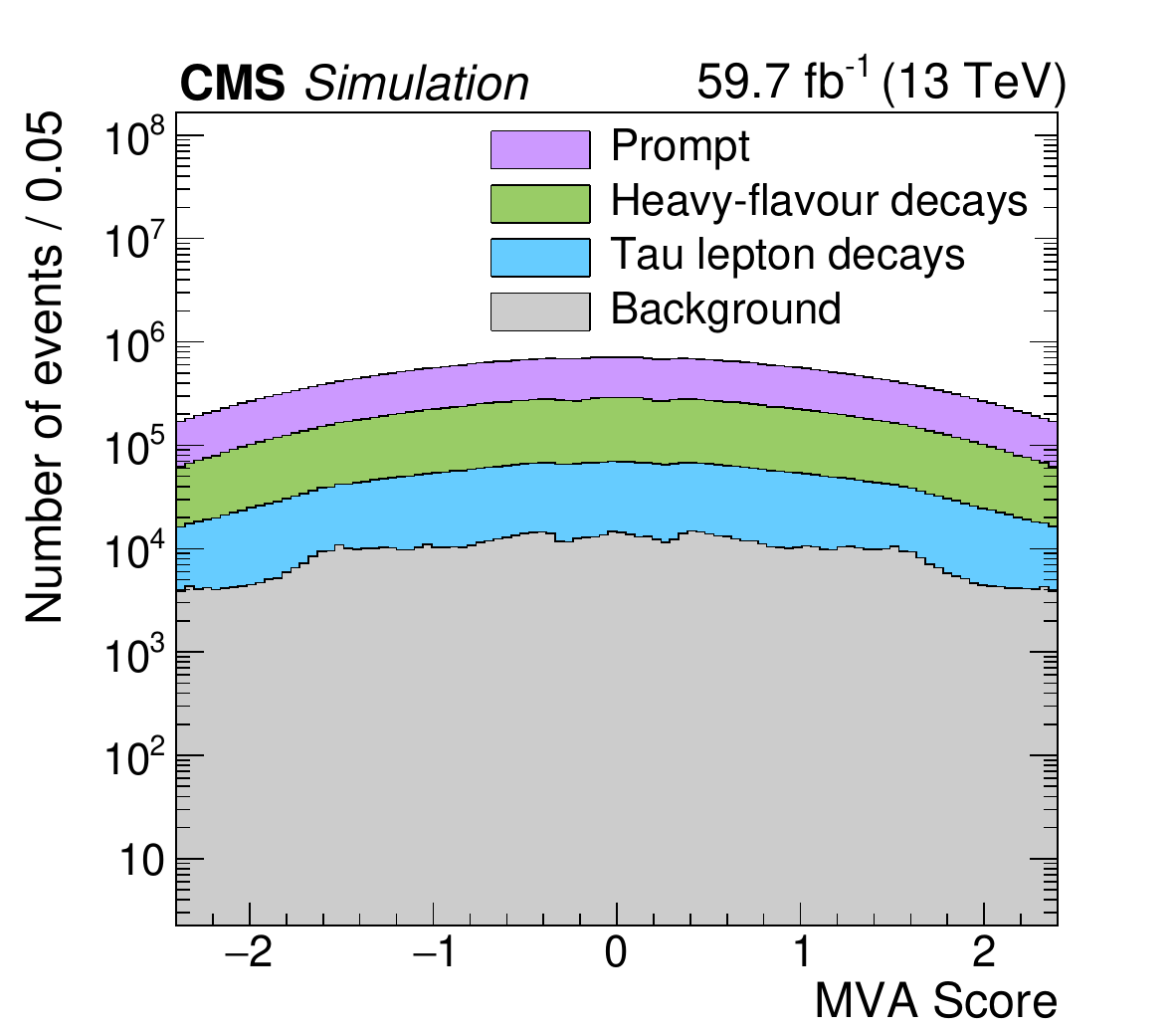}
  \caption{Composition of the simulated \ttbar sample used for training after muon preselection in terms of muon origin according to generator-level information. The composition is shown as a function of \pt (left) and $\eta$ (right). The last bin on the left figure includes events with $\pt > 100\GeV$.}
  \label{fig:pt_eta_stack}
\end{figure}

The sources of the muons included in the training sample are shown as a function of \pt and $\eta$ in Fig.~\ref{fig:pt_eta_stack}. As expected, the various signal contributions, as well as the background, exhibit different \pt profiles. The low-\pt region is enriched in muons from heavy-flavour hadron decays and background sources, whereas the higher \pt region presents a larger proportion of muons originating from prompt decays.

\begin{figure}[!ht]
  \centering
  \includegraphics[width=0.32\textwidth]{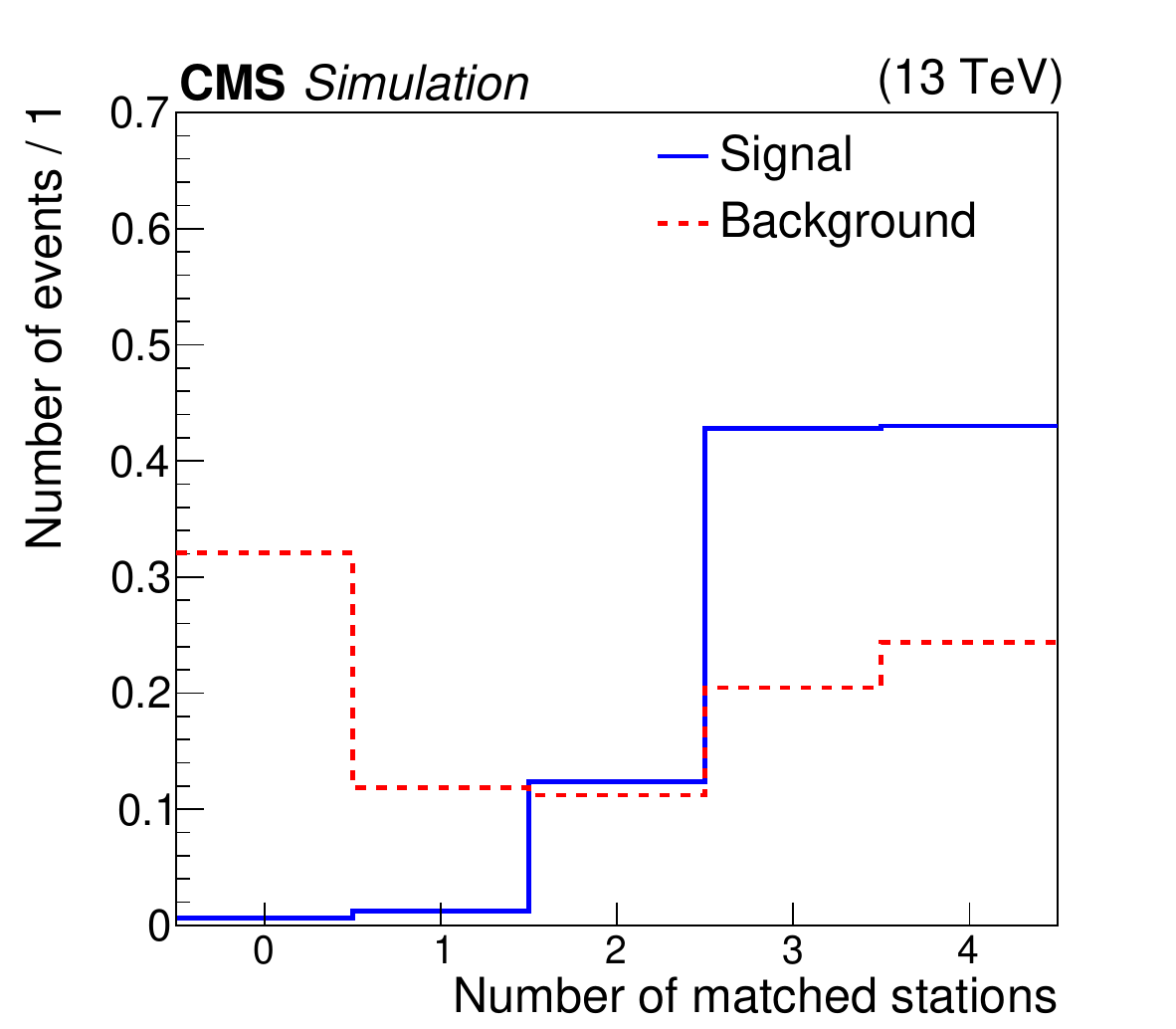}
  \includegraphics[width=0.32\textwidth]{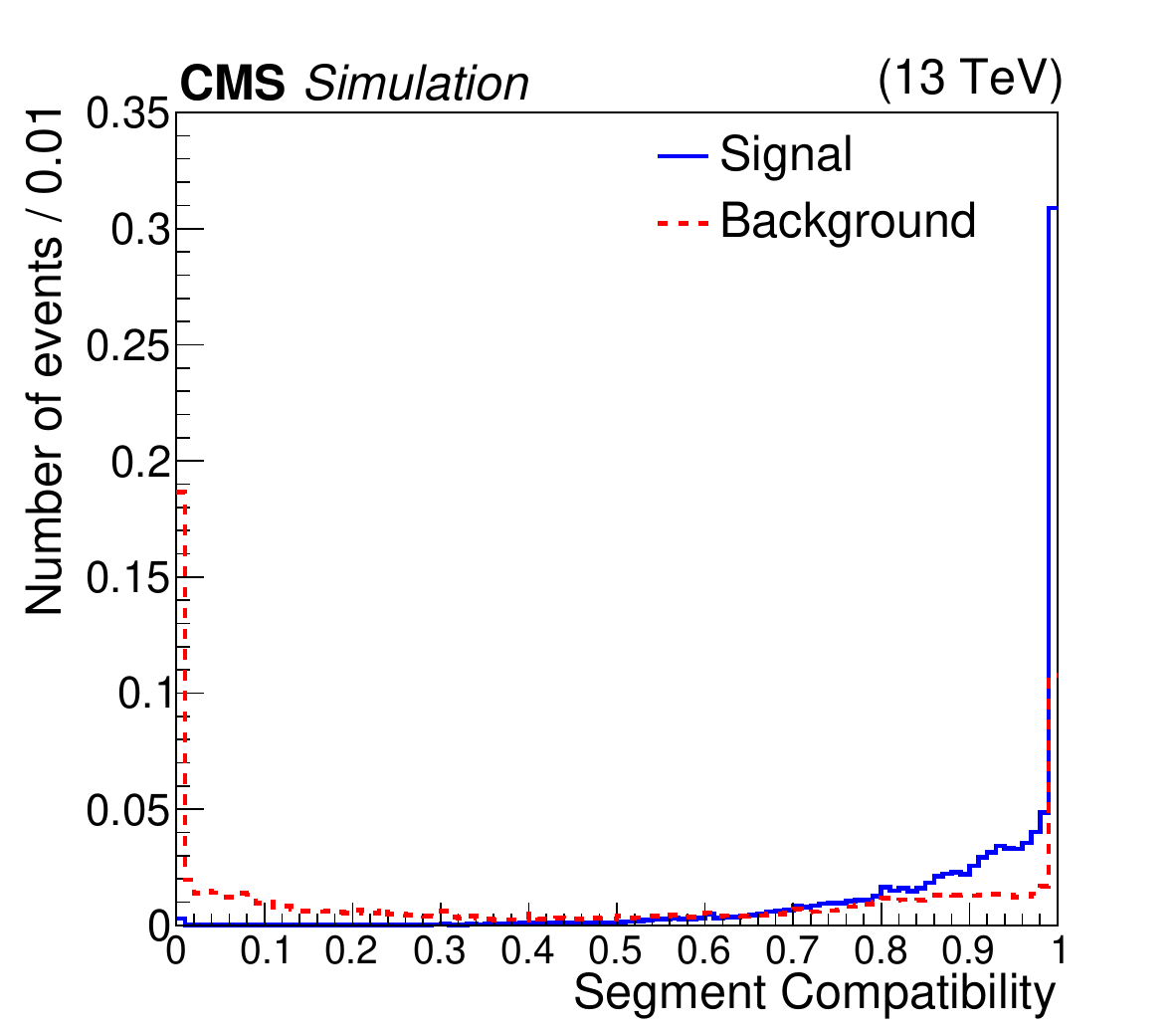}
  \includegraphics[width=0.32\textwidth]{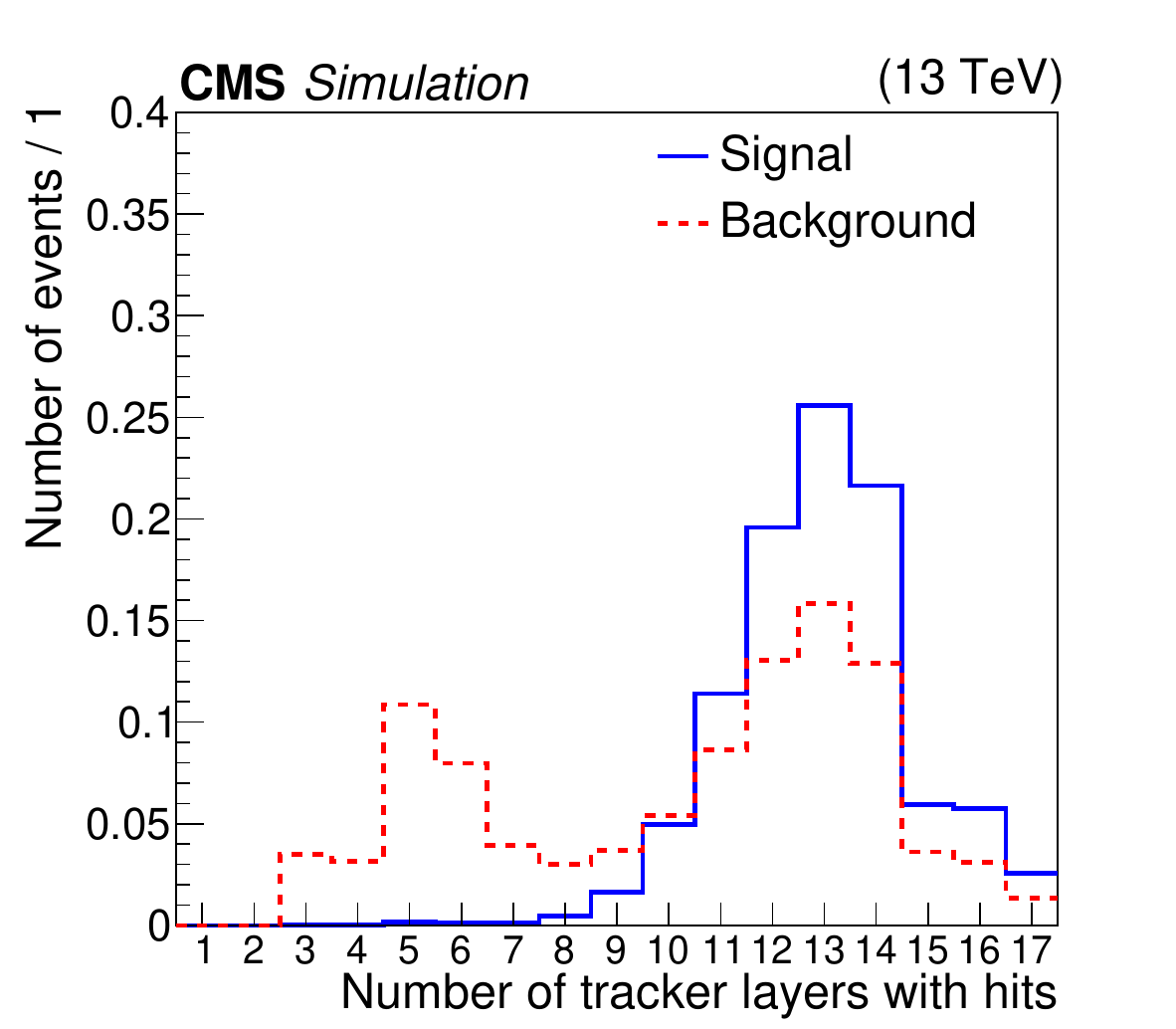} \\
  \includegraphics[width=0.32\textwidth]{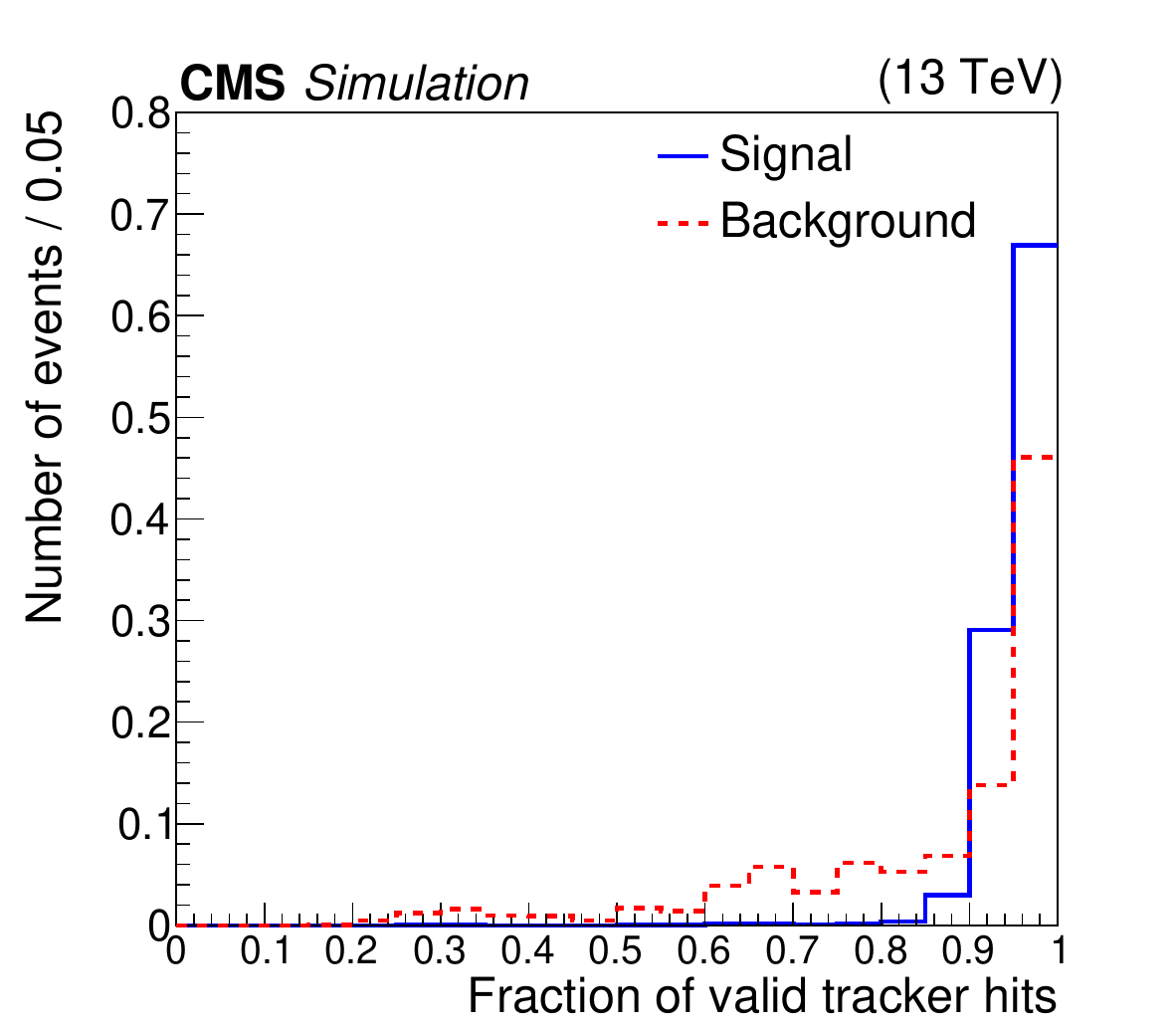}
  \includegraphics[width=0.32\textwidth]{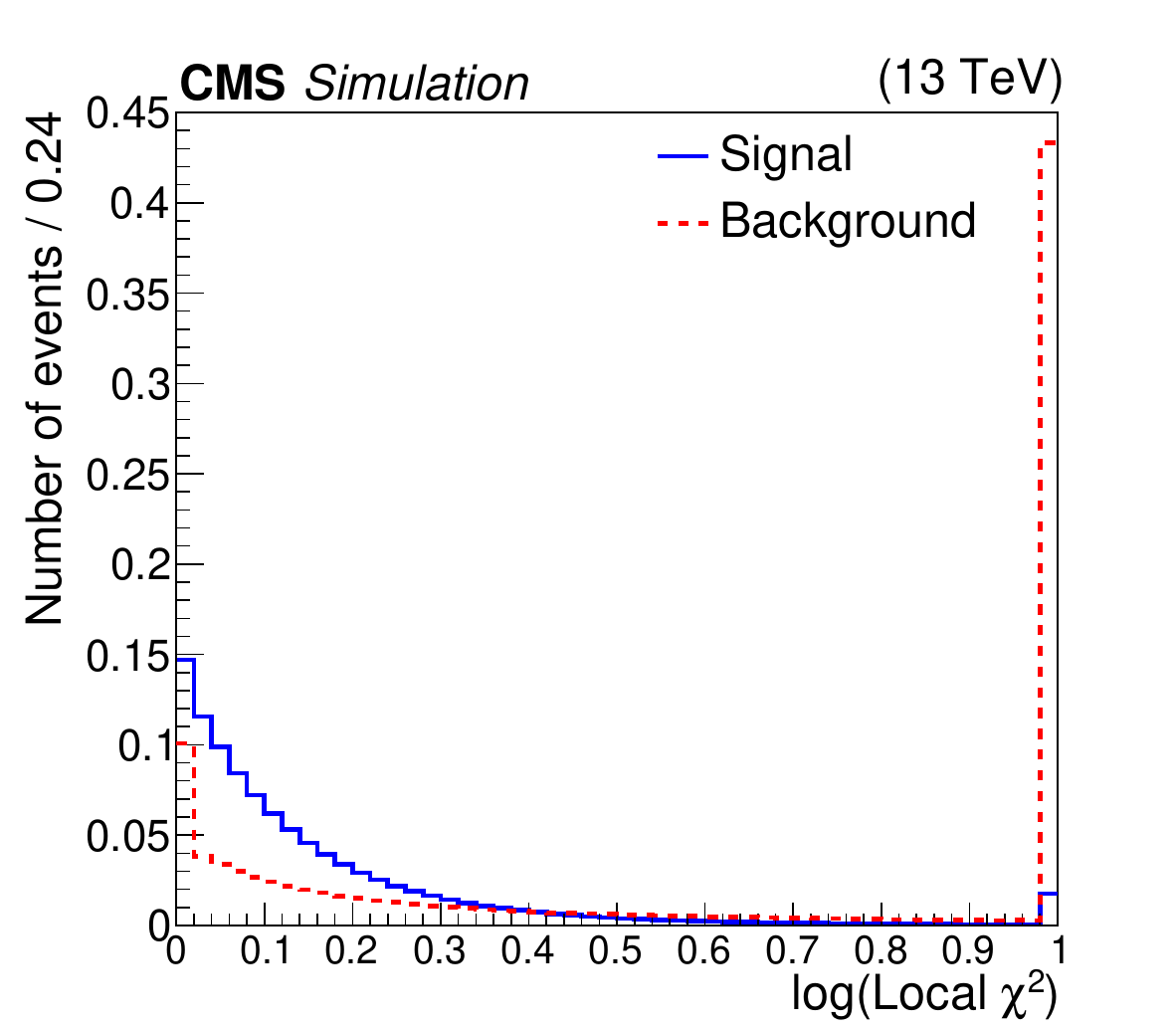}
  \includegraphics[width=0.32\textwidth]{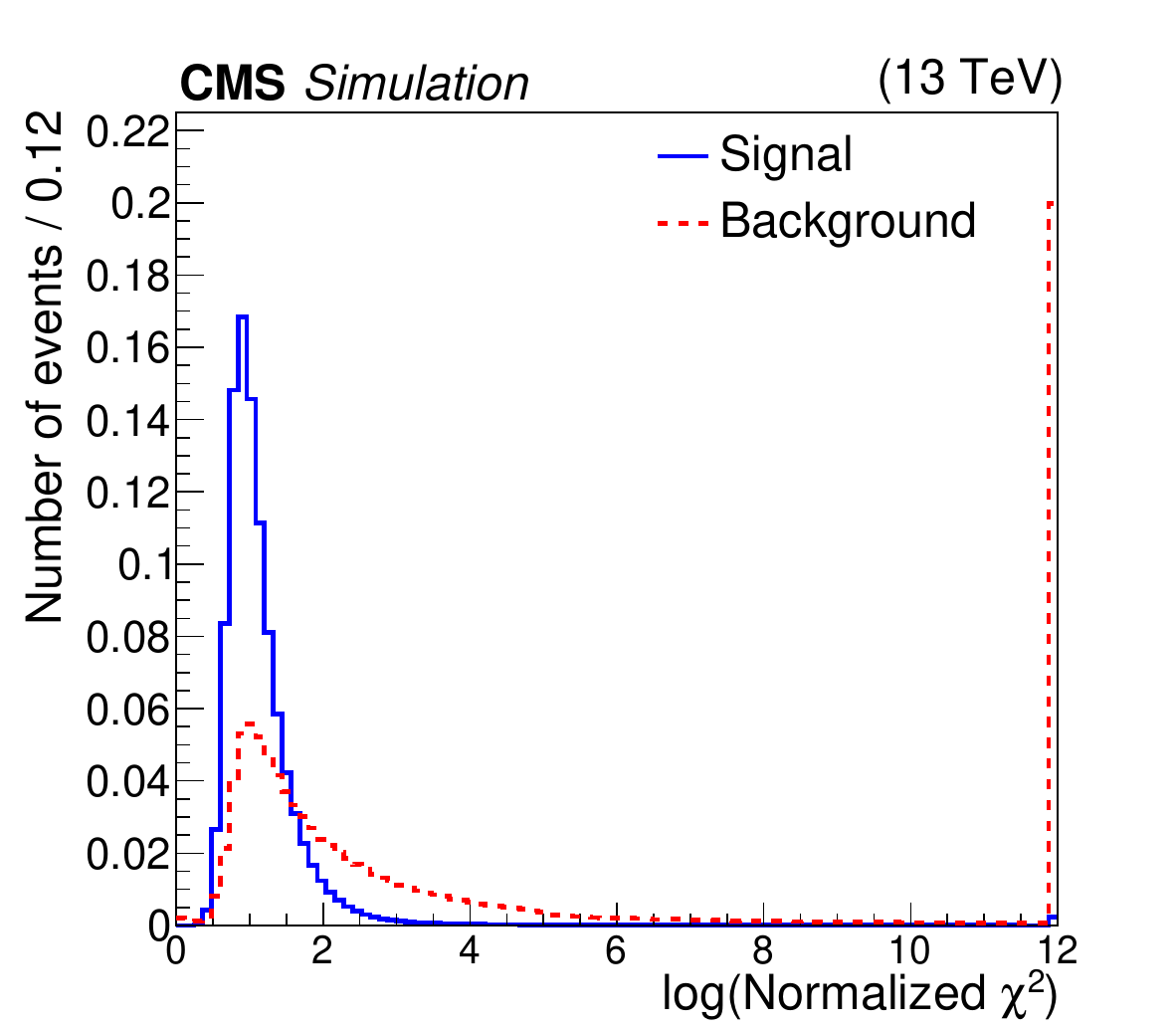} \\
  \includegraphics[width=0.32\textwidth]{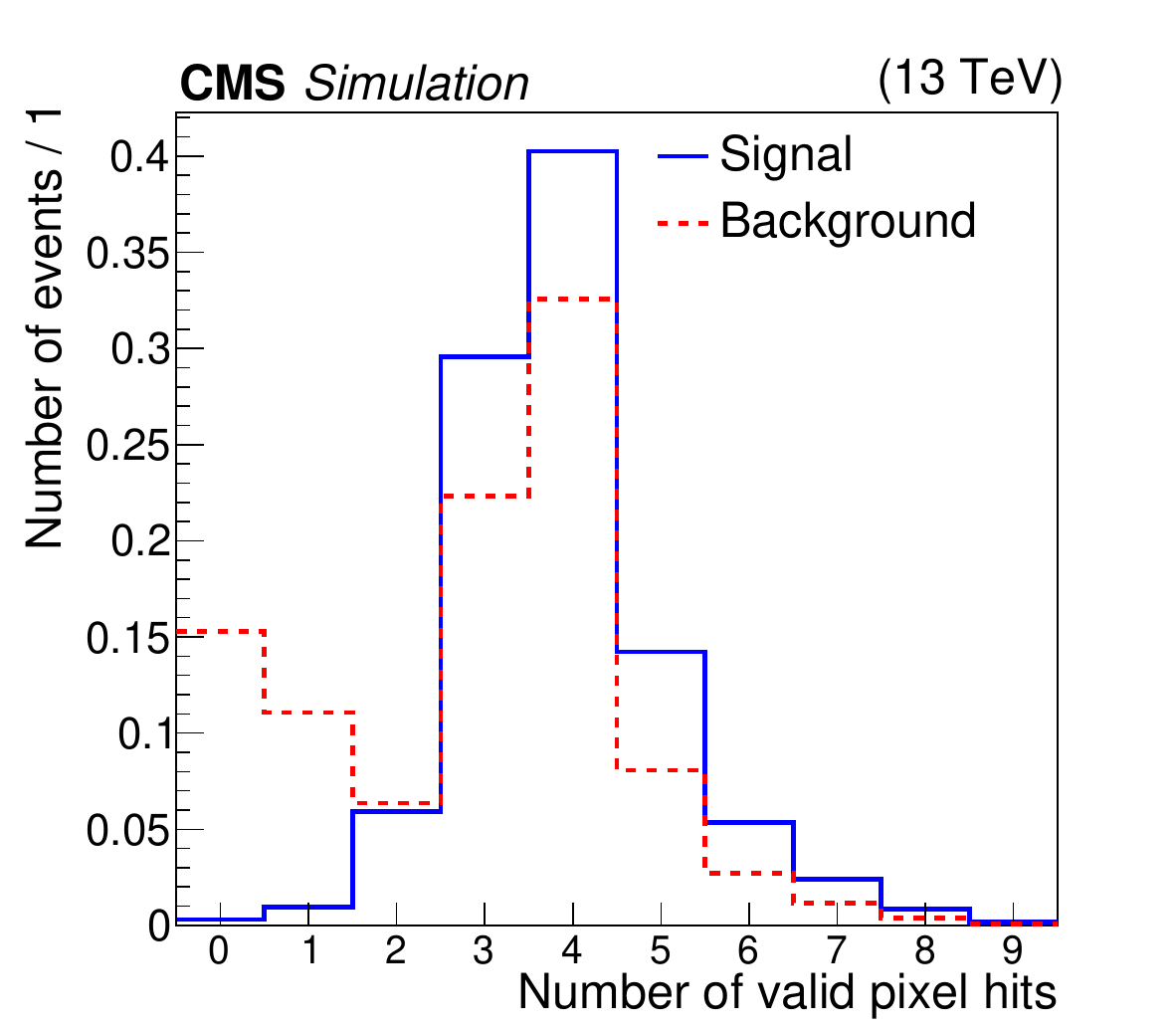}
  \includegraphics[width=0.32\textwidth]{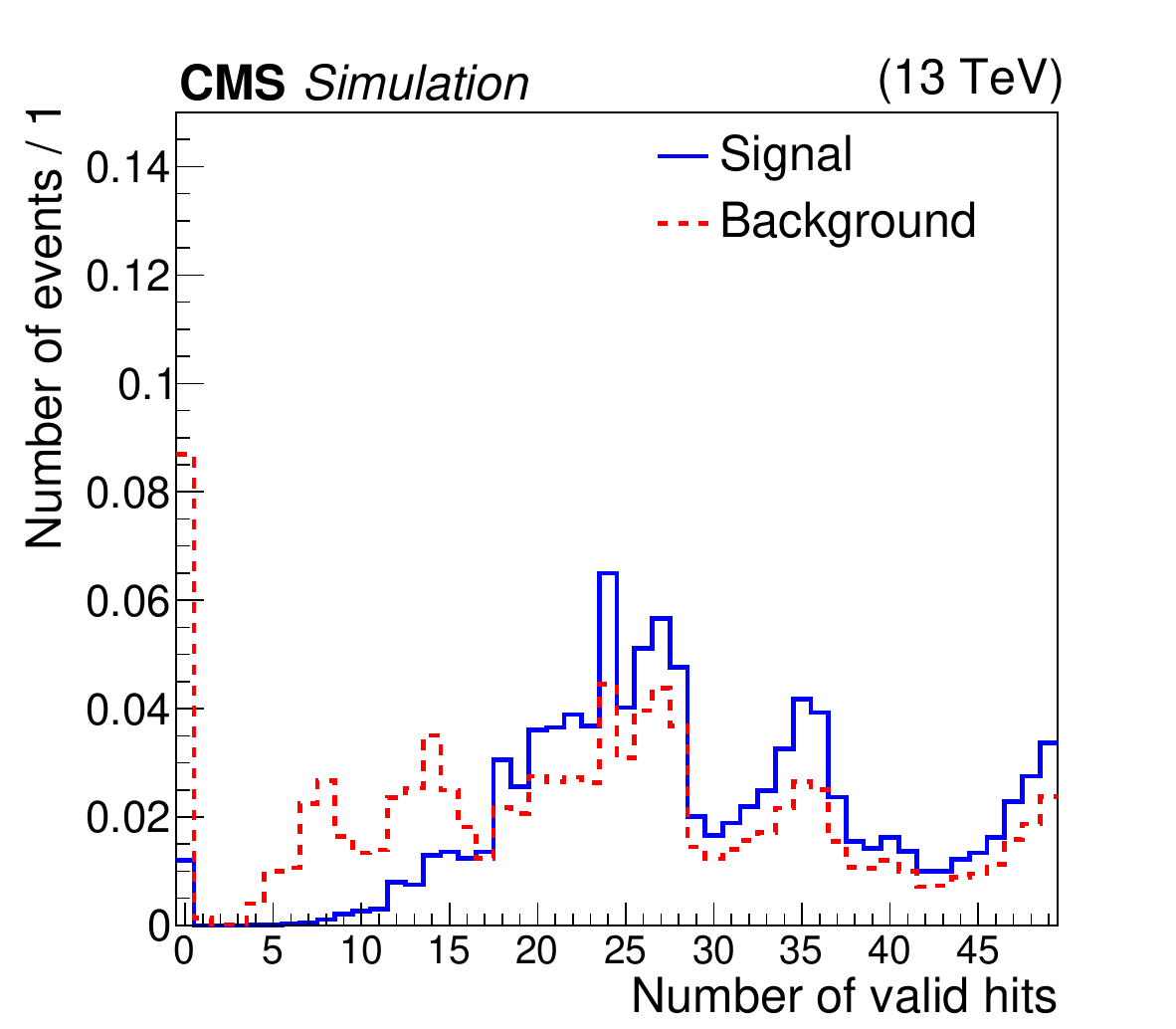}
  \includegraphics[width=0.32\textwidth]{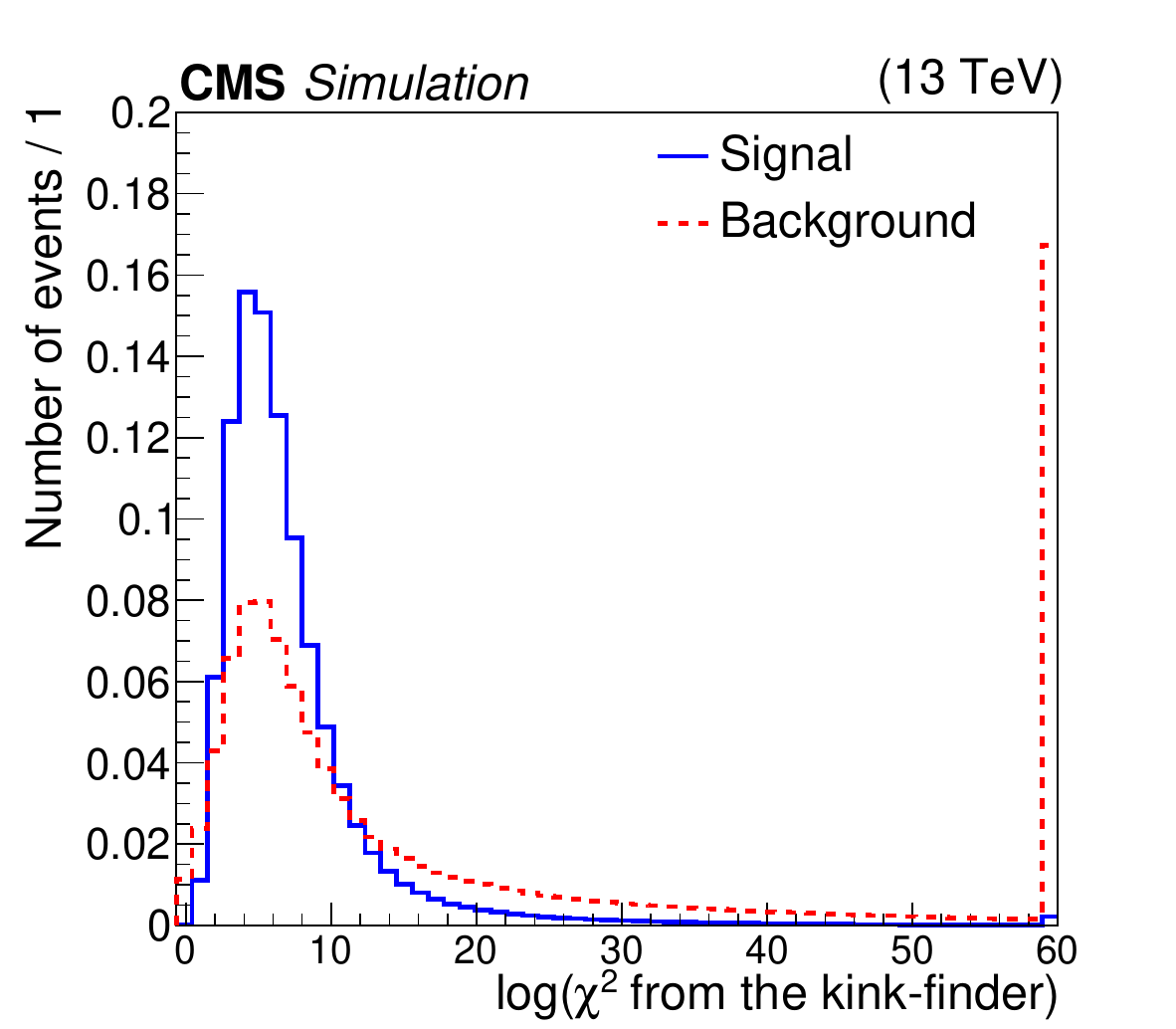}
  \caption{Distribution in the simulated \ttbar training sample of the number of matched stations (upper left), the segment compatibility (upper central), the number of tracker layers with hits (upper right), the fraction of valid tracker hits (middle left), the inner-standalone matching (middle central) and normalized \chisq of the muon fit (middle right), the number of valid pixel hits (lower left) and the total number of valid muon chamber hits (lower central), and the \chisq of the kink-finder algorithm (lower right), shown for signal and background, as defined in Section~\ref{sec:mvaid}. The last bin of each distribution contains the overflow events.}
  \label{fig:muonMVA_inputVariables}
\end{figure}

The training of the MVA is performed with all variables used in the standard cut-based identification criteria, with the exception of the compatibility with the PV, quantified with the muon impact parameter in the $xy$-plane and along the $z$-direction. These variables are excluded as we aim to obtain a discriminator to select both prompt and nonprompt muons produced in heavy-flavour hadron decays. To discard nonprompt muons, further isolation requirements or a custom impact parameter selection criteria should be applied at the analysis level. The input variables used are:

\begin{cmslist}
    \item The \pt and $\eta$ of the muon.
    \item ``Global muon'' flag.
    \item Normalized \chisq of the muon track fit.
    \item Number of muon stations with hits included in the muon track fit.
    \item \chisq of the kink-finder algorithm on the inner track defined in Ref.~\cite{CMS:2018rym}. The algorithm splits the inner track in two parts in several places along the trajectory and compares them. A large \chisq indicates that the two parts are not compatible with being a single track.
    \item \chisq of the position matching between the inner and standalone tracks (local \chisq). If there is no standalone track, it is set to 0.
    \item Segment compatibility: the compatibility of track segments in the muon system with the pattern expected for a minimum ionizing particle.
    \item Number of hits in the pixel detector used to fit the muon track.
    \item Number of tracker layers used in the muon track fit.
    \item Fraction of tracker hits used for the fit of the inner track.
    \item Number of matched stations: number of segments (one per station) reconstructed in the muon chambers and used in the global muon track fit.
\end{cmslist}
The \pt and $\eta$ bins are weighted to have the same distributions in both signal and background samples to avoid any kinematic bias in the muon classification. The distributions of the input variables used in the \ttbar training sample for signal and background muons passing the preselection requirements are shown in Fig.~\ref{fig:muonMVA_inputVariables}. For the three input variables defined as a \chisq the logarithm of the variable is presented.

A random forest~\cite{randomforest} classifier is trained using the \textsc{Scikit-learn}  0.19.1 package~\cite{scikit-learn}. All the parameters of the model (hyperparameters) are optimized, combining manual and grid search strategies, with the aim of achieving the best performance while
preventing overfitting, and hence reaching the highest possible accuracy on the classification. The optimal values of the hyperparameters
are 200 trees with a maximum depth of 8. The optimization was
performed using the area under the receiver operator characteristic (ROC) curve, as well as the efficiency and the background rejection.
The memory usage was also taken into account in the optimization, so slightly higher performance was traded for lower memory consumption.
The main challenge of this classification problem is the class imbalance, since the simulation contains many more signal than background muons.
The most important input variables in the training are the \chisq of the kink-finder algorithm on the inner track and the \chisq of the position matching between the inner and standalone tracks.

\begin{figure}[!ht]
\centering
\includegraphics[width=0.45\textwidth]{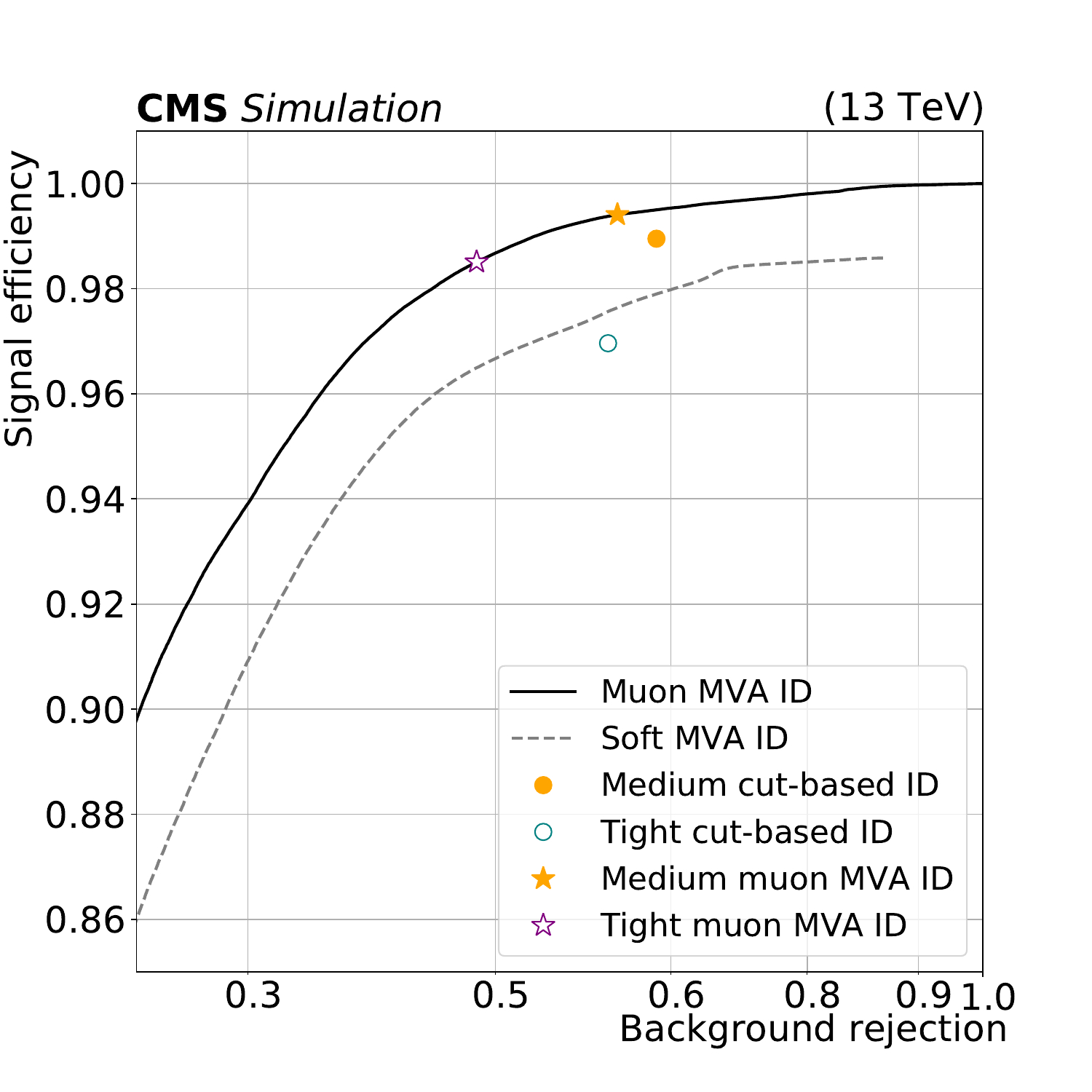}
\caption{The ROC curve for muons with $\pt > 10\GeV$ for the developed general muon MVA ID discriminator (black solid line) with the selected medium and tight WPs shown as orange solid and purple open stars, respectively. Orange solid and blue open points show the medium and tight WPs of the cut-based ID. The ROC curve of the soft MVA ID~\cite{CMS:2019bbr} is also shown (grey dashed line).}
\label{fig:probroc}
\end{figure}

The trained muon MVA ID discriminator offers a continuous variable, which provides flexibility to select the most suitable trade-off between signal and background efficiencies, known as working point (WP), for each analysis. Additionally, two WPs, medium and tight, are defined to provide a simple way to see the performance of the MVA in data. They provide an alternative discriminator for high-efficiency WPs while not necessarily as relaxed as the medium cut-based identification. The medium WP is defined as the value yielding the same background rejection of the cut-based selection medium WP, calculated for muons with $\pt > 20\GeV$ in the \ttbar simulated sample used to check the performance. A higher threshold in the muon MVA ID discriminator defines the tight WP to further reduce the background contamination by 10\% with respect to the medium WP, while still providing high efficiency. The defined thresholds correspond to MVA${}>0.08$ and 0.20 for the medium and tight WPs, respectively. Figure~\ref{fig:probroc} shows the ROC curve of the muon MVA ID with the selected muon MVA ID WPs. The cut-based medium and tight WPs, as well as the soft MVA ID~\cite{CMS:2019bbr} ROC curve are also shown for reference. The rates shown are computed selecting muons with $\pt > 10\GeV$ in the \ttbar sample that is not used for training. This plot summarizes the performance for all the available muon IDs.

\subsection{Prompt-muon identification and isolation}
\label{sec:leptonmva}

Muons originating from prompt decays of \PW, \PZ, or Higgs bosons, or from \PGt decays, appear isolated and pointing to the PV. This information is used to build an MVA discriminator that drastically reduces the contribution of nonprompt leptons to the selected muon sample. We refer to this discriminator as ``prompt-muon MVA''.
In addition to the minimal selection criteria detailed at the beginning of the section, muons used in the training are required to have a
mini-isolation smaller than 40\% of the \pt of the muon. In addition, the muon impact parameter with respect to the PV is required to be smaller than 0.05 cm in the $xy$-plane and 0.1 cm in the $z$-direction.
The three-dimensional impact parameter of the muon track with respect to the PV, divided by its uncertainty, which corresponds to its
significance ($d/\sigma_d$) is required to be smaller than 8.
These requirements reject muons that are very unlikely to have a prompt origin, while bounding some of the MVA input variables, enhancing its performance.
The signal is defined as the muons selected by these criteria and matched to a generator-level muon produced in the prompt decay of the particles mentioned above, whereas all the muons not fulfilling these criteria are considered as background. The sample of signal (background) muons is obtained from simulated samples of \ttH (\ttbar) events.

The input variables used in the training include properties of the muons, information regarding the muon isolation, and, if present, properties of the jet associated with the muon. To be considered as associated, the jet must contain the muon candidate.
The training variables are:
\begin{cmslist}
    \item Muon \pt and \abseta.
    \item Charged component of the mini-isolation variable, defined as $I_{\PGm}^{\text{charged}} = \sum_{\text{charged}} \pt / \ptmu$.
    \item Neutral component of the mini-isolation variable, corrected for PU effects with the effective areas method~\cite{CMS:2018rym}, defined as $I_{\PGm}^{\text{neutrals}} = \max \big( 0, \sum_{\mathrm{neutrals}} \pt - \rho \, \mathcal{A} \, (R/0.3)^2 \big) / \ptmu$.
    \item Muon-to-jet \pt ratio variable, $\pt^{\text{ratio}}$:
    the ratio of the transverse momentum of the muon to the transverse momentum of the nearest jet, $\ptmu/\pt^{\text{jet}}$. If no jet associated to the muon is present, this variable is set to $1 / (1+\Irel)$, where \Irel is the relative isolation.
    \item Muon relative \pt variable, \ptrel: the component of the muon momentum in direction transverse to the jet, $\ptrel = \pmu \sin\theta$,
    where $\theta$ denotes the angle between the muon and jet momentum vectors and \pmu, the magnitude of the muon momentum. If no jet associated with the muon is present, this variable is set to zero.
    \item Jet \PQb tagging score:
    the value of the \DeepJet \PQb tagging algorithm discriminator~\cite{CMS-DP-2023-005} of the associated jet. When such a jet is not present, the variable is set to zero.
    \item Jet charged constituents:
    the number, $N_{\text{charged}}$, of charged PF candidates within the associated jet. Tracks associated with those particles must be within $R < 0.4$ of the muon and are required to come from the PV to enter the counting. Minimal track quality, \pt, and impact parameter criteria are also applied. If such a jet does not exist this variable is set to zero.
    \item Longitudinal ($d_z$) and transverse ($d_{xy}$) impact parameters. Since these variables span over several orders of magnitude, their logarithmic value, $\log\big( d_{z(xy)} / 1\cm\big)$, is used.
    \item Significance of the three-dimensional impact parameter.
    \item Muon segment compatibility variable.
\end{cmslist}

\begin{figure}[!ht]
  \centering
  \includegraphics[width=0.32\textwidth]{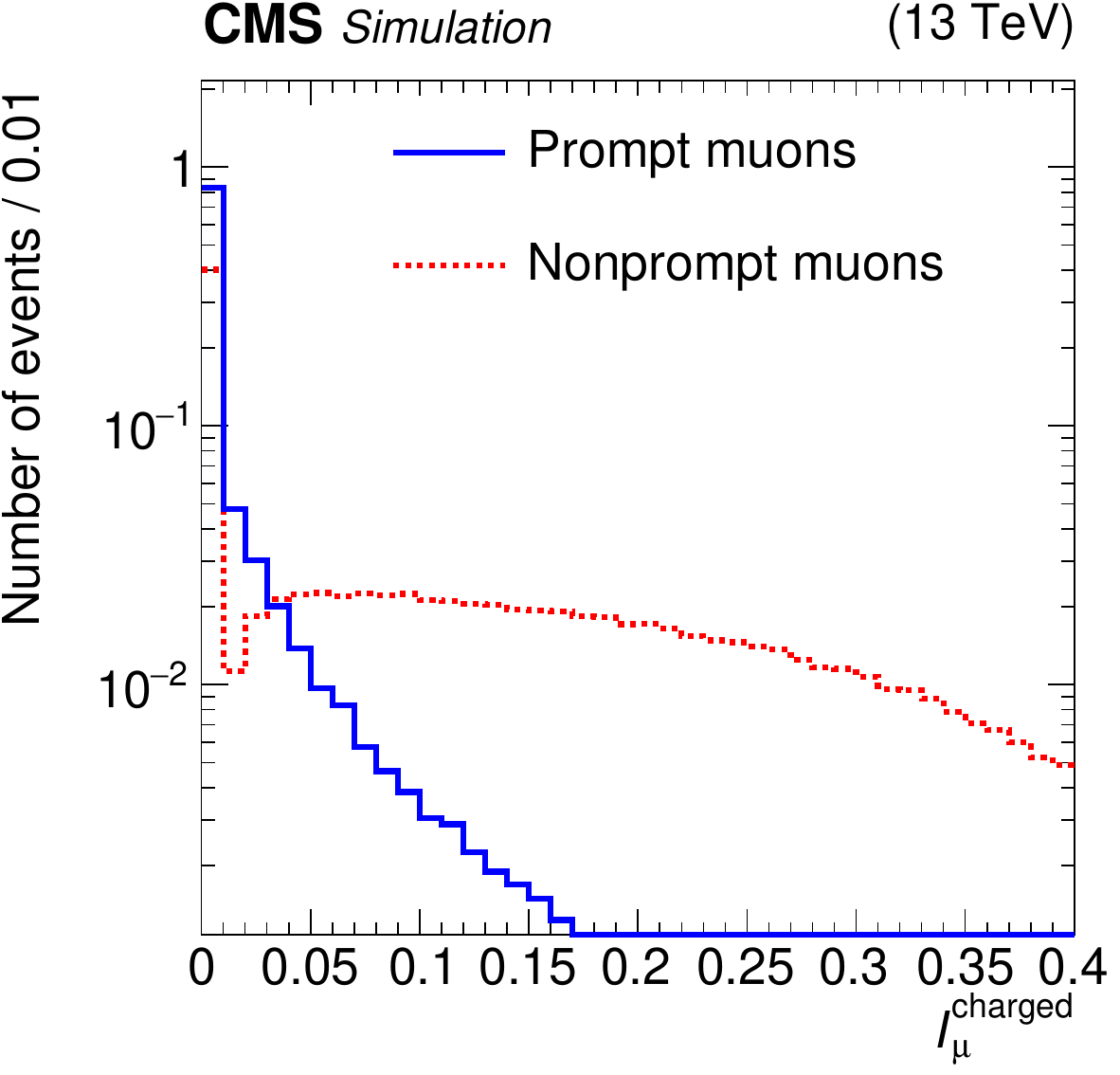}
  \includegraphics[width=0.32\textwidth]{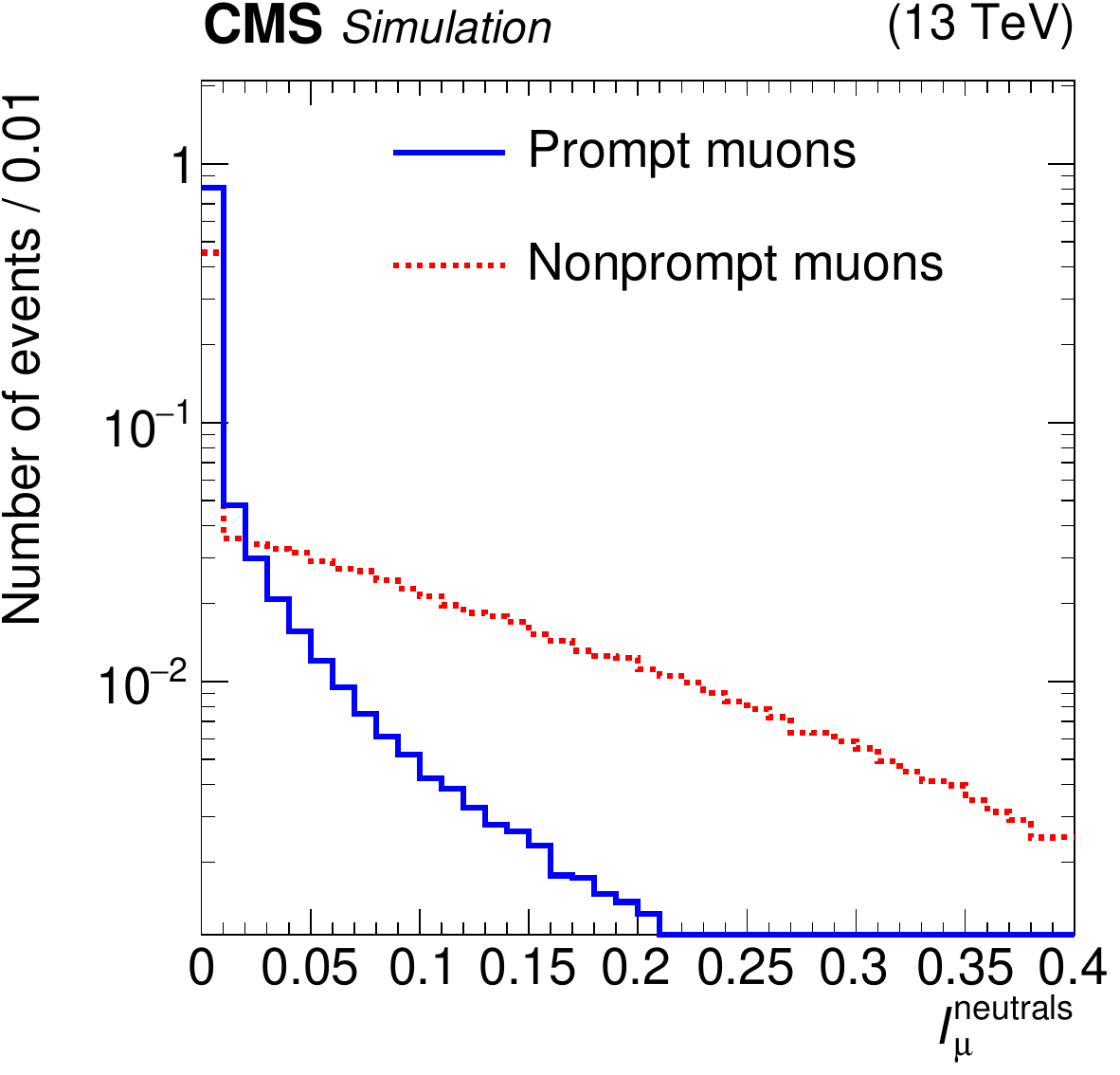}
  \includegraphics[width=0.32\textwidth]{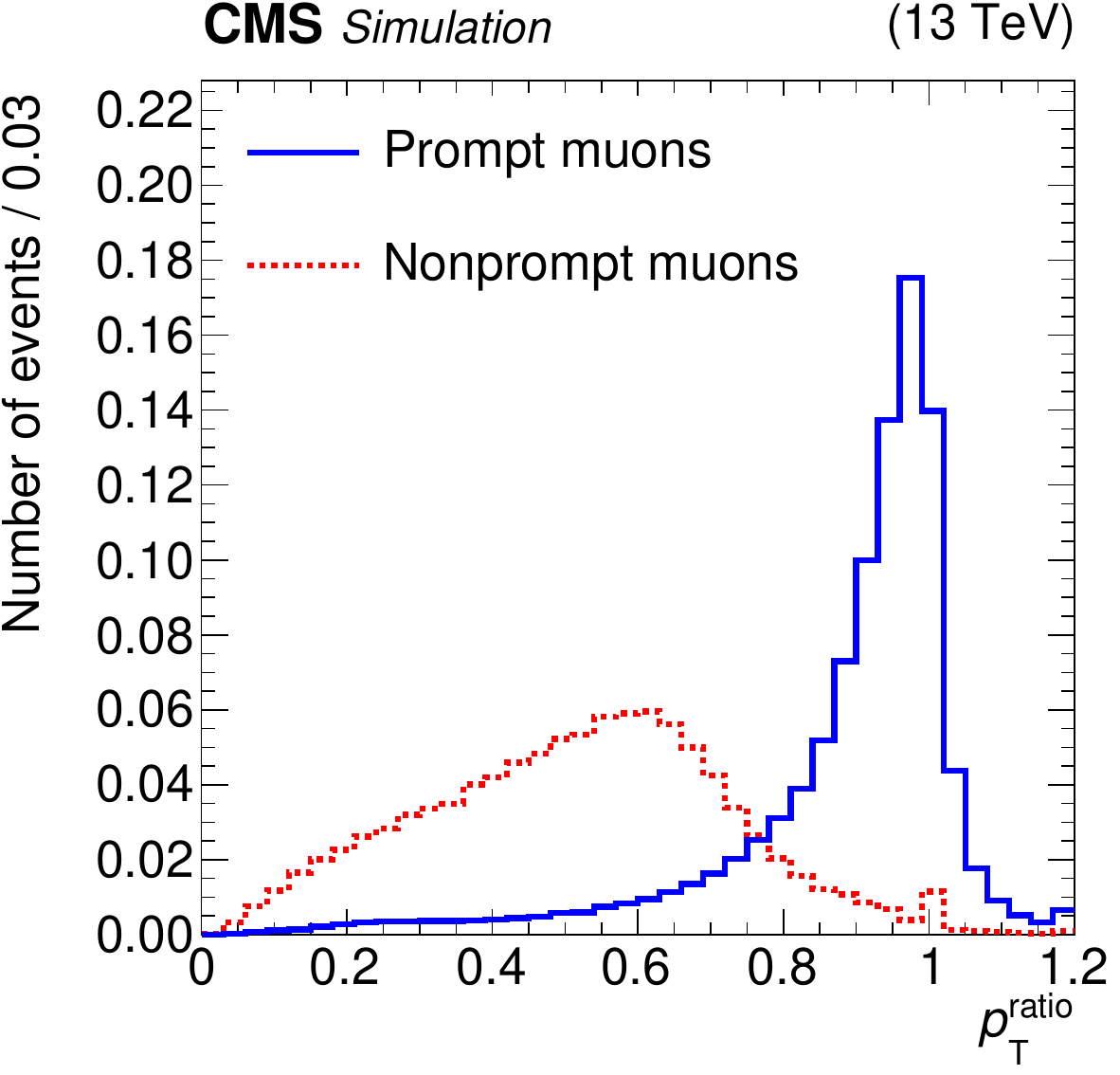} \\
  \includegraphics[width=0.32\textwidth]{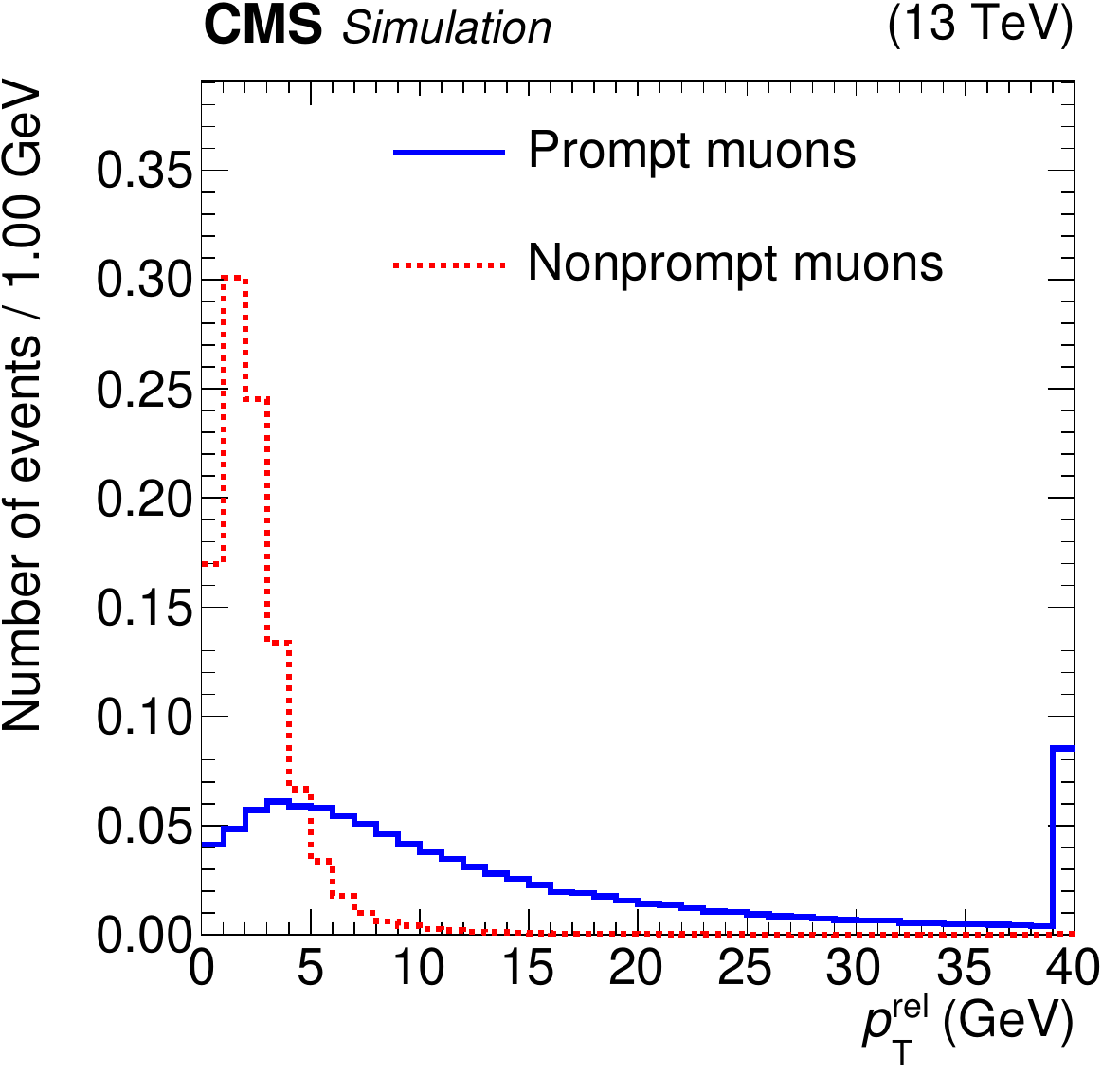}
  \includegraphics[width=0.32\textwidth]{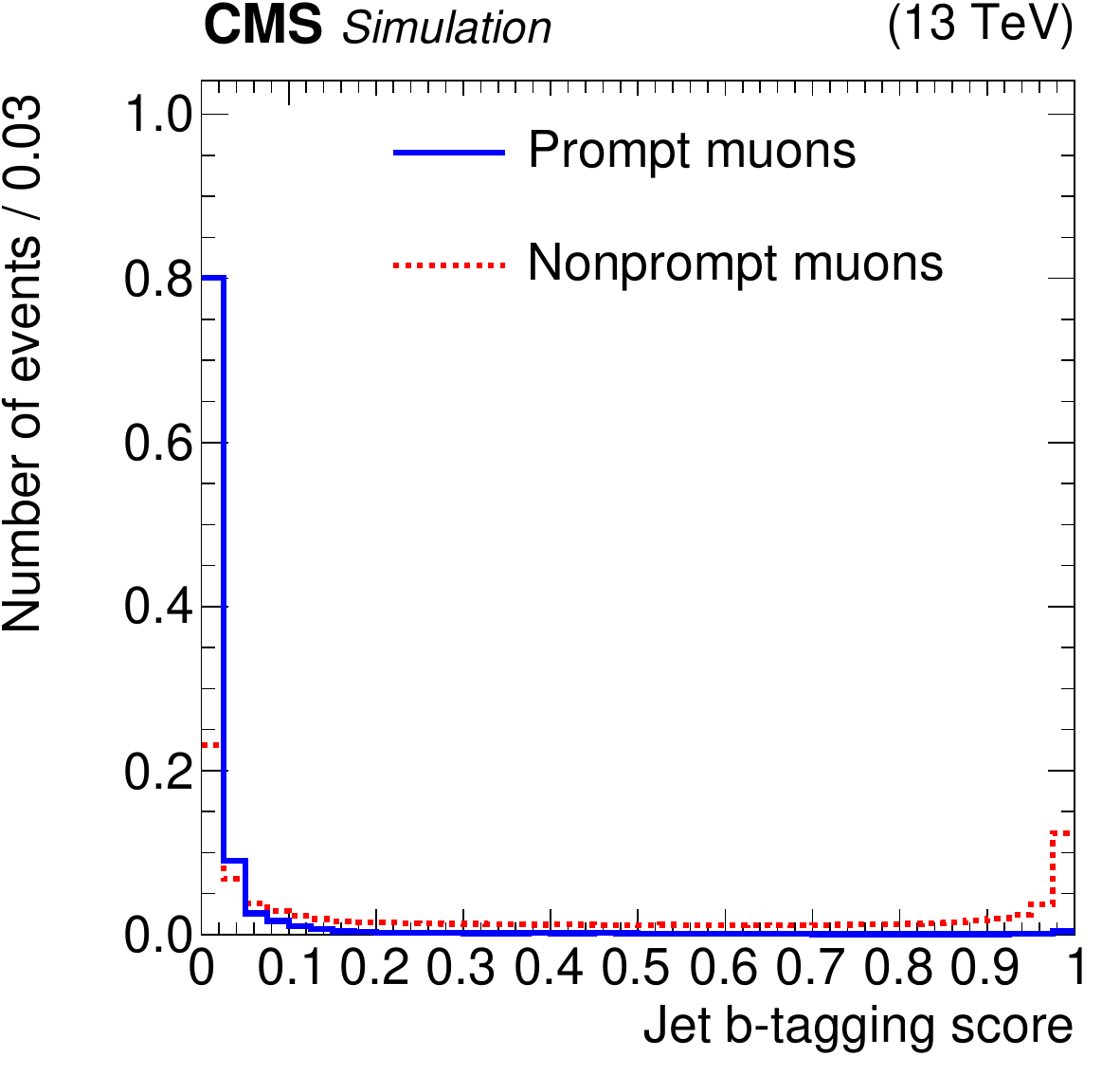}
  \includegraphics[width=0.32\textwidth]{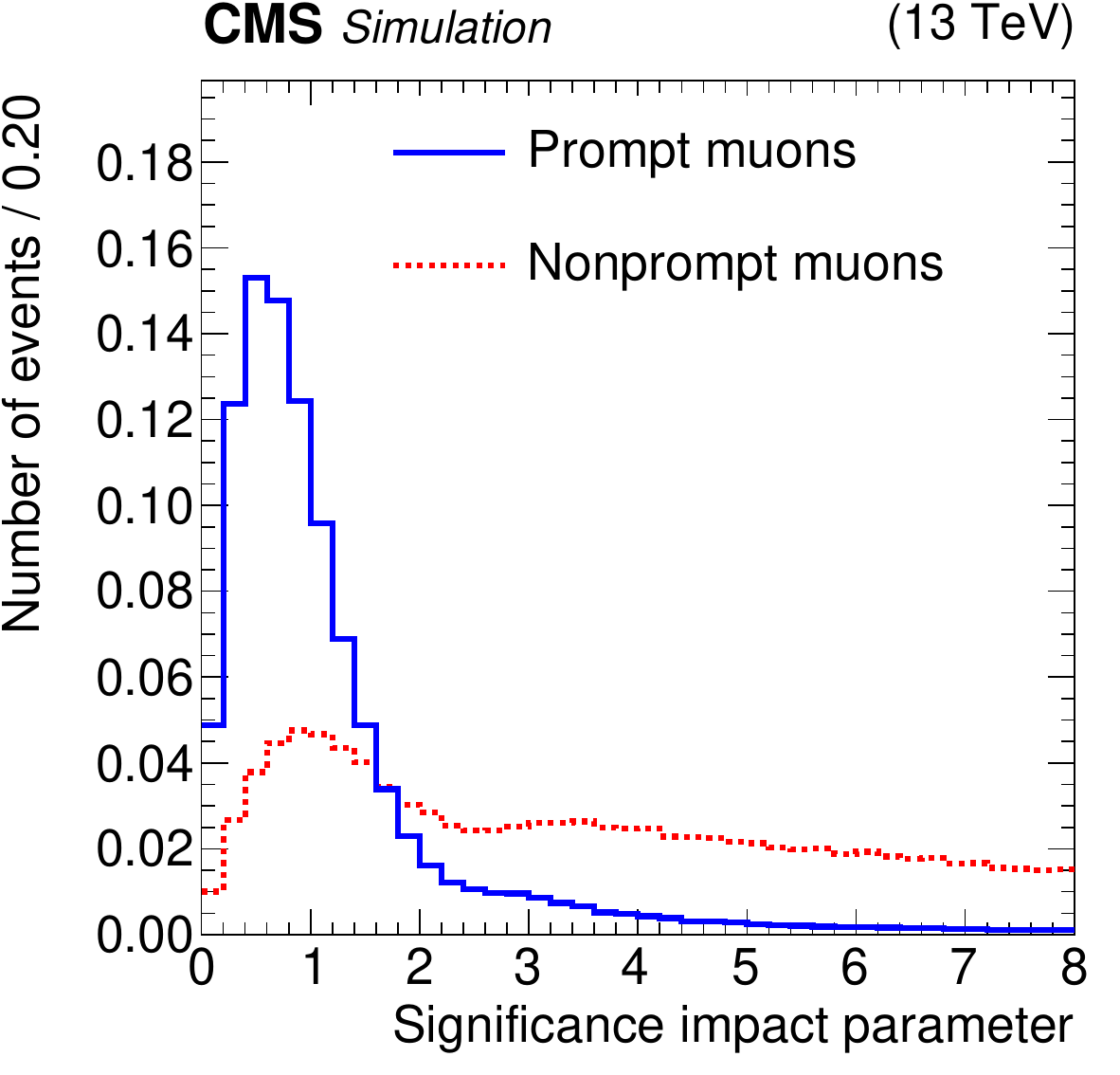} \\
  \includegraphics[width=0.32\textwidth]{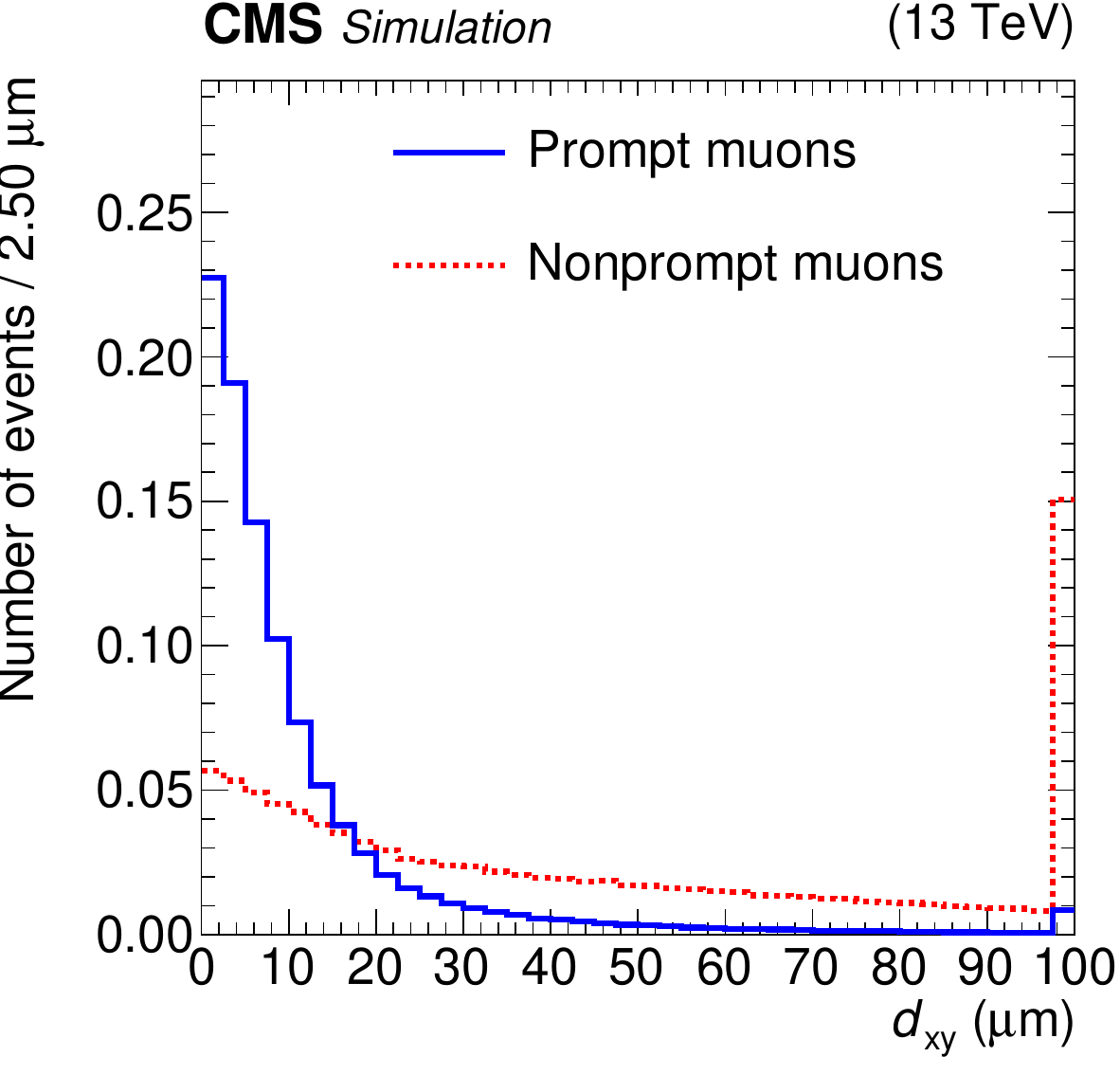}
  \includegraphics[width=0.32\textwidth]{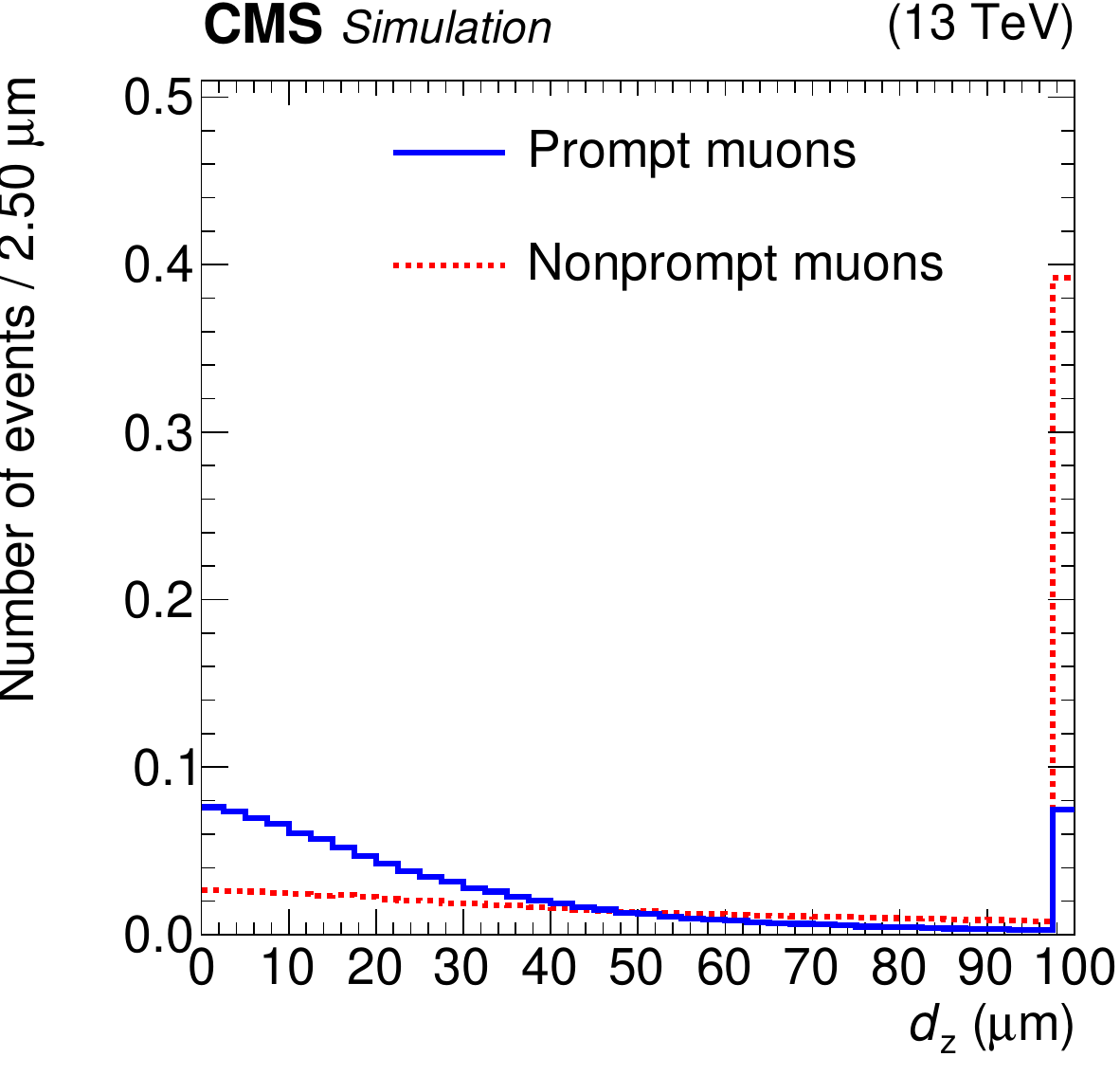}
  \includegraphics[width=0.32\textwidth]{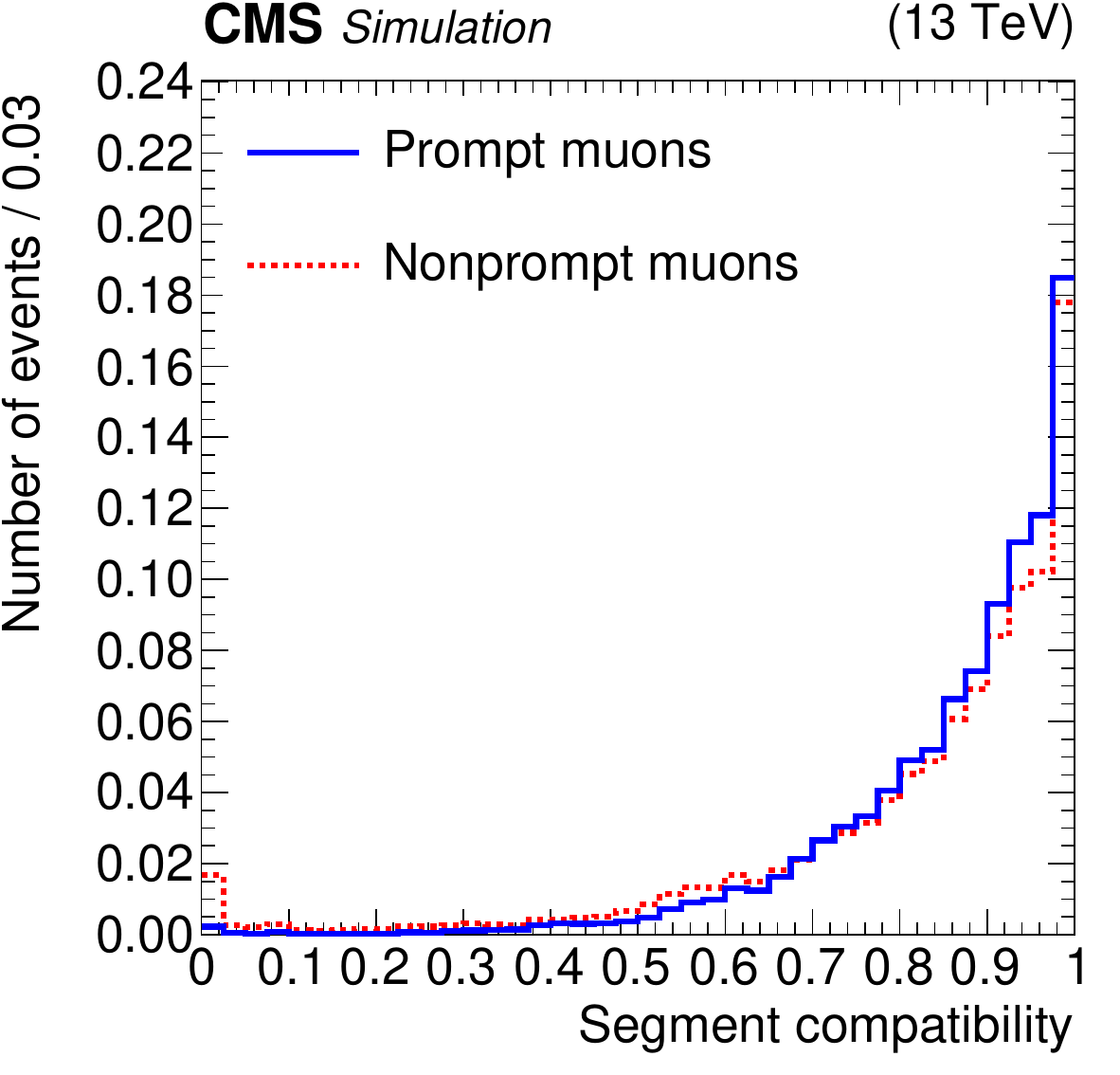}
  \caption{Simulated distributions of the charged component (upper left) and the  neutral component (upper central) of mini-isolation, the muon to jet \pt ratio (upper right), the jet relative \pt (middle left), the score of the associated \DeepJet \PQb tagging algorithm (middle central), the significance of the impact parameter (middle right), the impact parameter in the transverse (lower left) and longitudinal (lower central) direction between the muon and the PV, and the segment compatibility (lower right) for prompt and nonprompt muons.}
  \label{fig:leptonMVA_inputVariables}
\end{figure}

The classifier is trained using the \textsc{TMVA} 4.2.1 implementation~\cite{Voss:2007jxm} of a boosted decision tree (BDT) with gradient boosting. The distributions of signal and background muons, as predicted by the simulations, are shown in Fig.~\ref{fig:leptonMVA_inputVariables} as functions of the input variables. The hyperparameters of the BDT have been chosen to achieve the best performance: the algorithm used is a gradient boost with 1000 trees of maxium depth of 4. This optimization was performed using the area under the ROC curve as metric, since no WP was predefined.

\begin{figure}[!ht]
\centering
\includegraphics[width=0.5\textwidth]{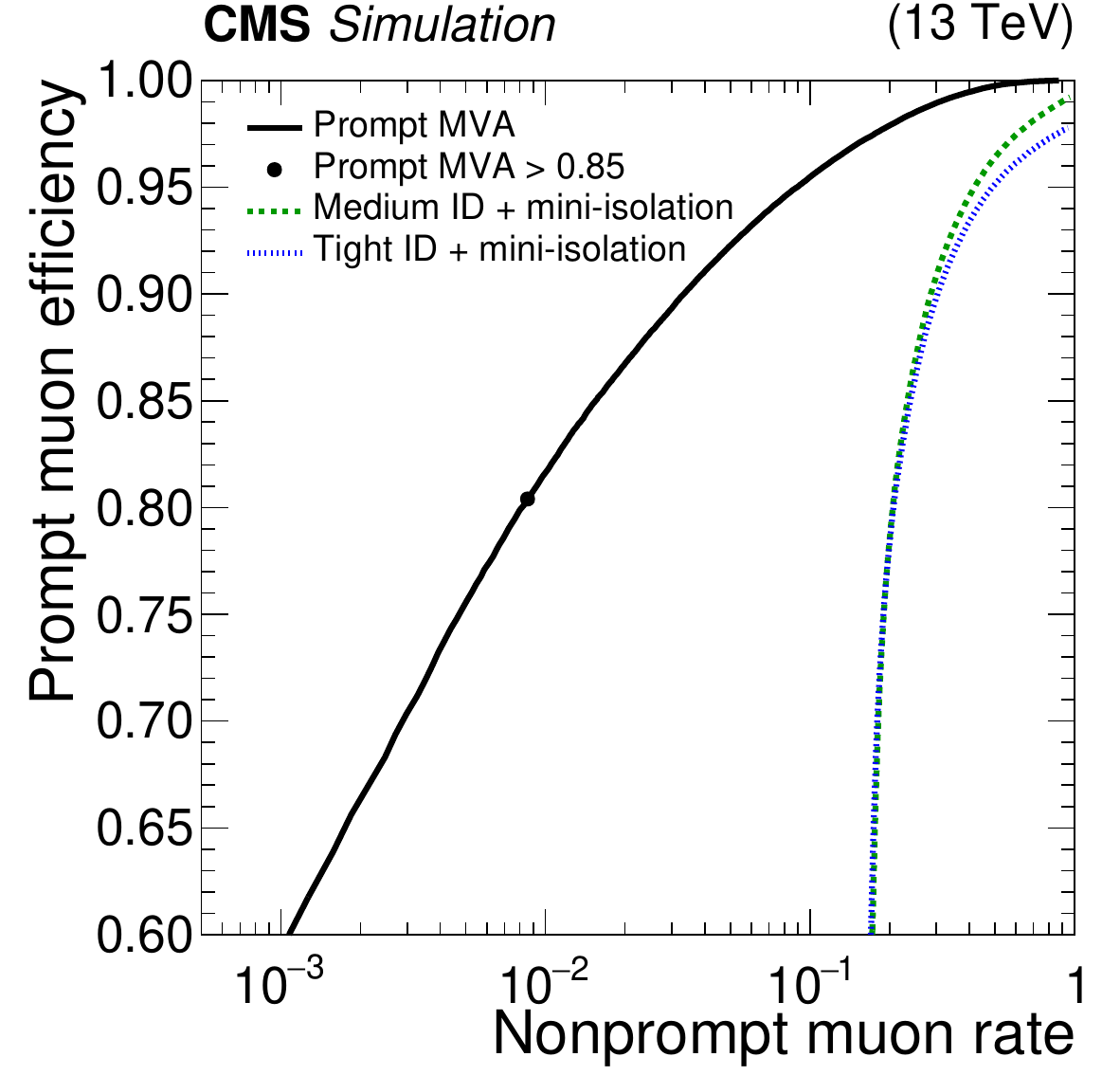}
\caption{ROC curve for the prompt-muon MVA, and for the tight and medium cut-based criteria together with requirements on mini-isolation. For a set of cuts on the different discriminators, the efficiency is shown as a function of proportion of nonprompt muons passing the WP selection.}
\label{fig:roc_leptonmva}
\end{figure}

The performance of the discriminator is shown in Fig.~\ref{fig:roc_leptonmva} together with the performance obtained using a cut-based approach, which combines a requirement on the tight or the medium cut-based identification WP with a range of requirements on the mini-isolation variable. For the prompt-muon MVA, a WP used already in Ref.~\cite{CMS:2020mpn} is defined to have a MVA score larger than 0.85, which gives about 80\% efficiency in the \ttH events and a nonprompt muon rate of the order of 6\ten{$-3$} in \ttbar events, where efficiency (nonprompt rate) is defined as the fraction of signal (background) muons passing the selection.

\section{Performance in data}
\label{sec:run2perf}

The performance of the discriminators is evaluated using both data and simulated events, and discussed in this section.

The efficiency is measured as a function of the muon \pt in two $\eta$ regions using the ``tag-and-probe method'', which selects muons coming from the \PZ boson decay, as described in Ref.~\cite{CMS:2010svw}. Tag muons are required to pass the tight cut-based identification criteria and, to avoid any bias in the efficiency measurement, to be geometrically matched with the muon that triggered the event. Probe muons are required to pass the loose cut-based identification criteria. Efficiency is defined as the fraction of probe muons that pass the MVA identification requirement.

In simulation, any background to this selection is removed by requiring the tag and the probe muons to be produced by the decay of a \PZ boson, looking at the generation-level information. In collision events, a parametric model of the dimuon invariant mass distribution with a background and a signal component is built. The signal is parameterized using the templates predicted by the simulated sample of DY events convolved with a Gaussian distribution to account for potential resolution differences between data and simulation. To model the background, the convolution of an exponential decay distribution with an error function is used. The exponential is used to model the distribution at high-mass values whereas the error function is used to model the low-mass regime.

The systematic uncertainties in the efficiencies estimated using the tag-and-probe method are smaller than 1\%~\cite{CMS:2018rym} and are correlated between the different discriminators.

The systematic uncertainties in the background contamination are strongly dependent on the topology of the background that contaminates each specific analysis; in particular, when the background contamination is due to muons with very different topology (\eg multijet events) with respect to those of the signal region. For the prompt-muon MVA, which uses \PQb tagging and other variables strongly dependent on the event topology, the uncertainties can be larger. However, no large dependences on the event topology have been observed at analysis level~\cite{CMS:2020mpn}. In either case, the analyses making use of these MVA selections should tailor the systematic uncertainty in their specific cases.

Several analyses in CMS have already used the prompt-muon MVA discriminator with an equivalent training to that described in this paper.

Some examples include the evidence of \ttH production~\cite{CMS:2020mpn,CMS:2018fdh,CMS:2018uxb}, in which the use of the prompt-muon MVA (including an equivalent training for electrons) leads to a reduction in the non\-prompt-lepton background from \ttbar events by a factor of 4 with respect to the use of an equivalent cut-based technique.
A second example is the measurement of the $\PW\PZ$ process~\cite{CMS:2021icx,CMS:2019efc}, or the multiboson production at $\sqrt{s}=5.02\TeV$~\cite{CMS:2021pqj}, where the use of this identification signficantly reduced the poorly modelled nonprompt-lepton background arising from \Zjets and \ttbar events.
A third example is the search for electroweak production of supersymmetric particles~\cite{CMS:2021cox} where the use of this discriminator improved the nonprompt-lepton background rejection, therefore limiting the associated systematic uncertainties.
Finally, an alternative training of this discriminator was also employed in the observation of $\PQt\PZ\PQq$ production~\cite{CMS:2018sgc} and four top quark production ($\ttbar\ttbar$)~\cite{CMS:2023tttt}, as well as in other precision top quark measurements~\cite{CMS:2022tkv, CMS:2019too, CMS:2021ugv}.

The muon MVA ID was not previously used in any CMS analysis, but it offers an alternative discriminator that could be used for high-efficiency WPs.
In addition, it features a weaker dependence on the number of PU vertices compared with the standard cut-based approaches, which may be useful in the context of changing data-taking conditions or high PU scenarios expected in \Run3. These points are of particular interest for those precision analyses that are not statistically limited, such as the top quark or the electroweak precision measurements, allowing for a nearly background-free signal region. In addition to reducing the amount of background, this may significantly reduce also its associated systematic uncertainties.

\subsection{Muon MVA ID performance}

The efficiency for each of the WPs is measured in both data and DY simulation using the tag-and-probe method.
The background contamination and the signal purity are measured in a \ttbar simulated sample that was not used for training. The background contamination is defined as the number of background muons passing the identification criteria divided by the total number of background muons that also fulfill the cut-based loose identification criteria, and the purity is calculated as the number of signal muons divided by the total number of muons, for all these muons passing the identification criteria. A slight improvement of the purity for muons with \pt between 10 and 20\GeV is observed when using the MVA discriminators, whereas purity values for higher \pt are similar to that of the cut-based ID.

\begin{figure}[!ht]
  \centering
  \includegraphics[width=0.32\textwidth]{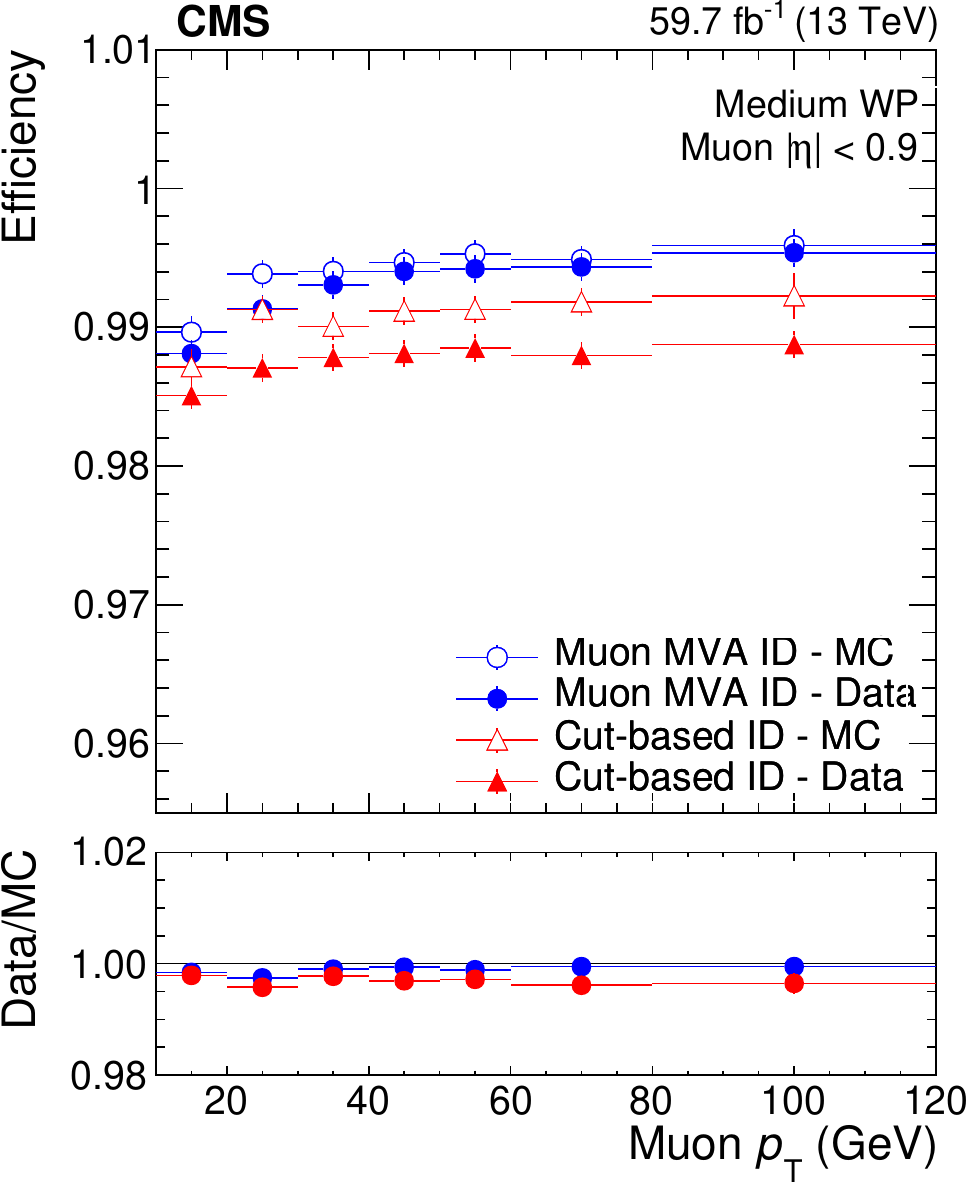}
  \includegraphics[width=0.32\textwidth]{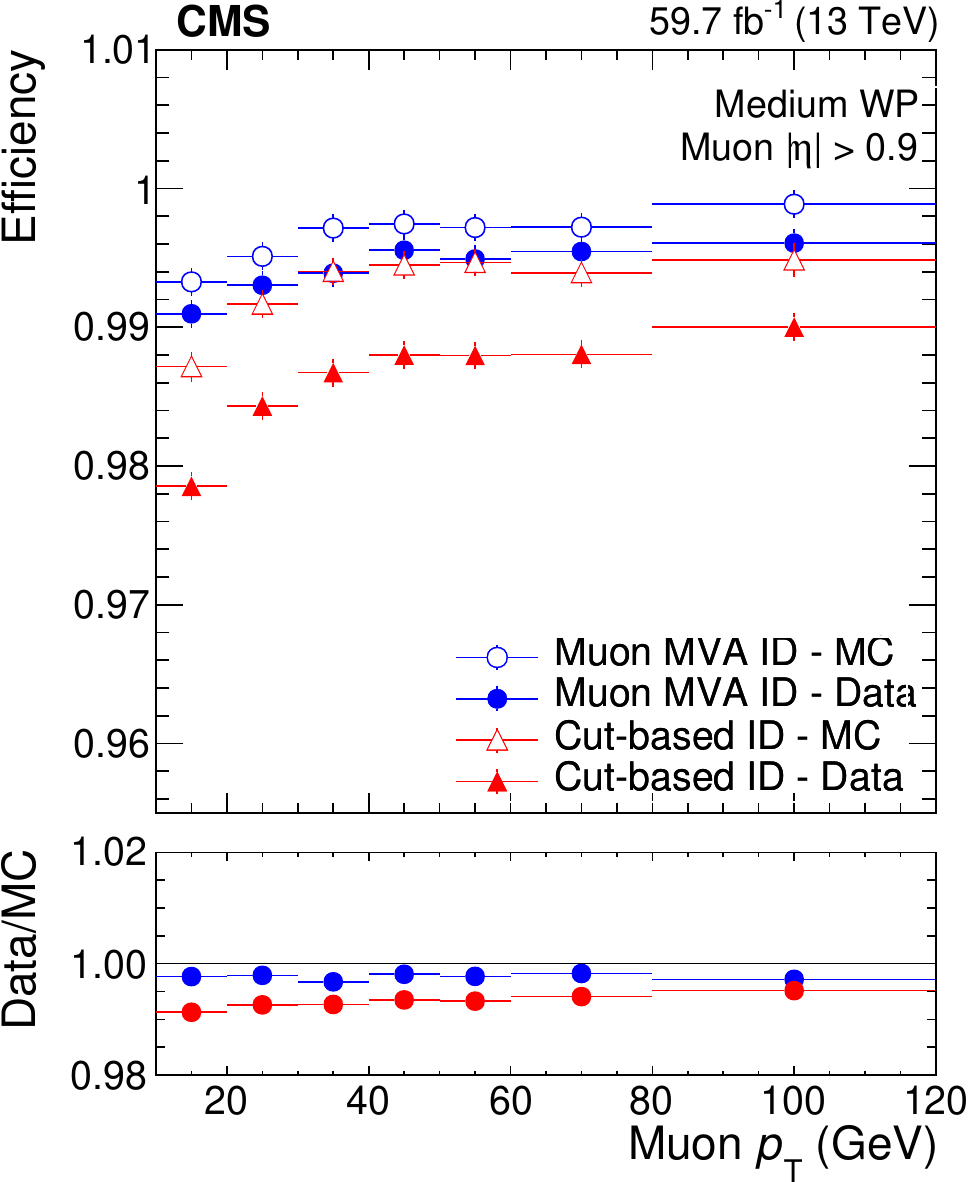} \\
  \includegraphics[width=0.32\textwidth]{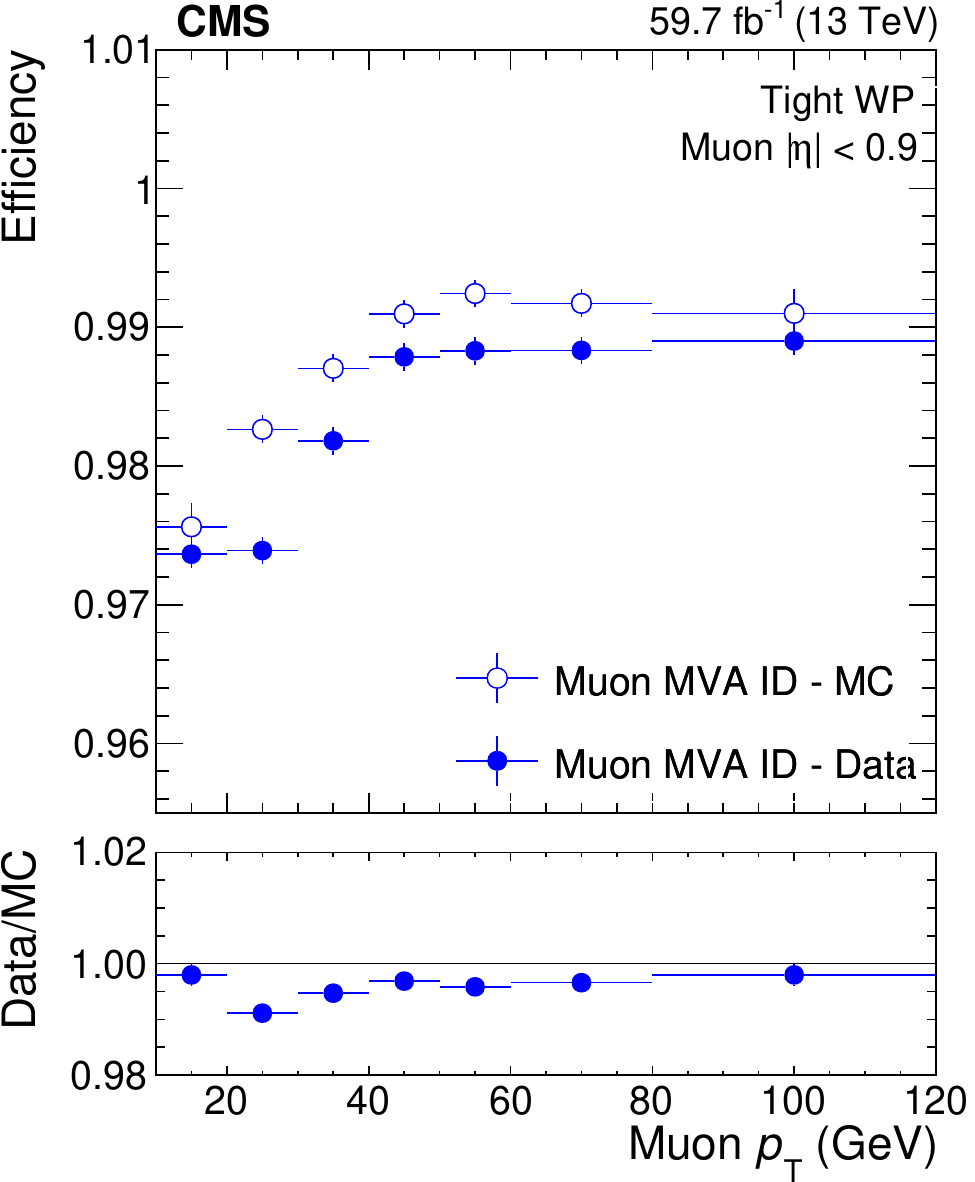}
  \includegraphics[width=0.32\textwidth]{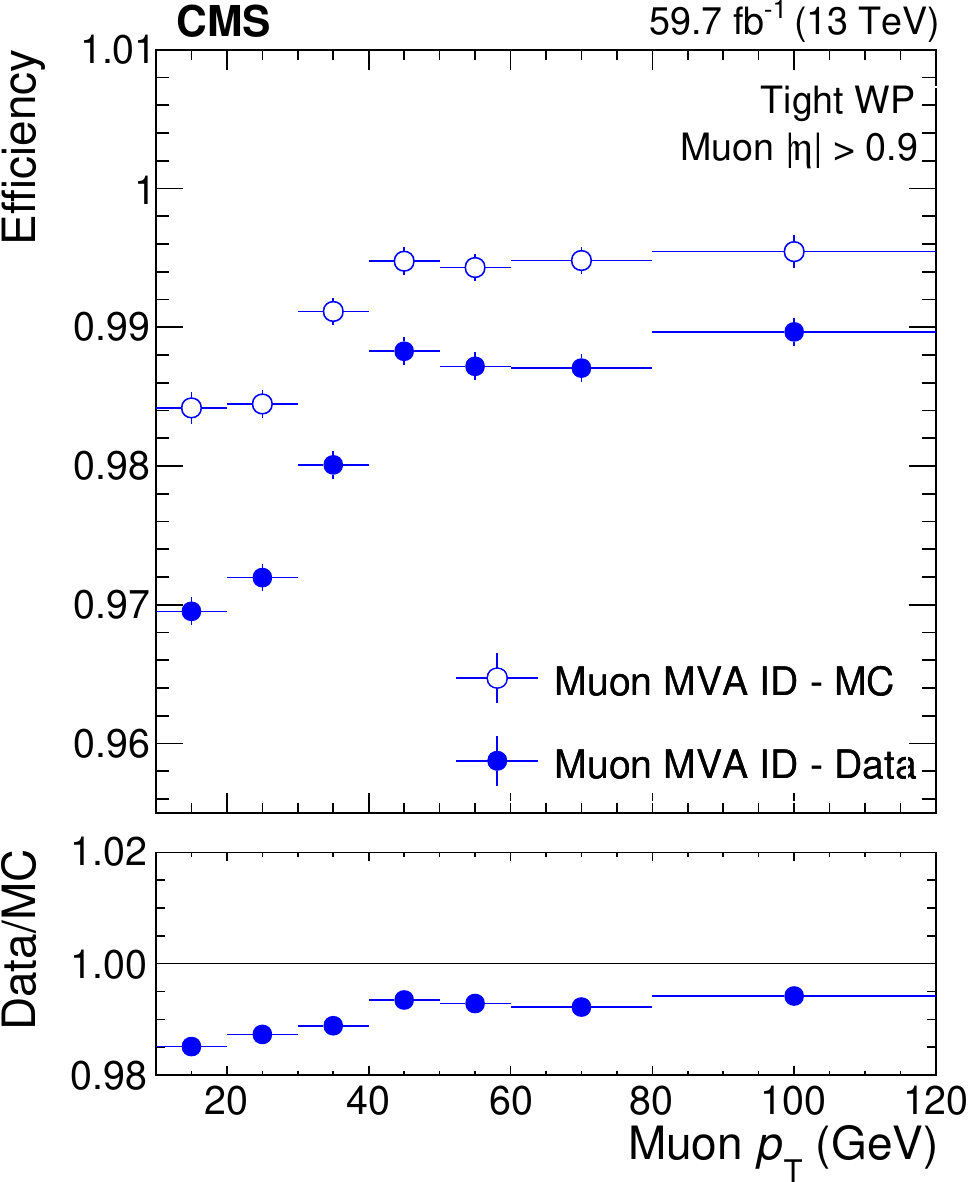}
  \caption{Muon identification efficiency for the medium (upper) and tight (lower) WPs as a function of \pt for muons with $\abseta < 0.9$ (left) and  $\abseta > 0.9$  (right). Blue dots show the muon MVA ID performance both for the 2018 data set and DY simulation, whereas red triangles show the efficiency of the medium cut-based ID used during \Run2. The data to MC ratio is also shown.  The efficiencies of the muon MVA ID are similar in both $\eta$ regions.}
  \label{fig:EffvsPt}
\end{figure}

\begin{figure}[!htp]
  \centering
  \includegraphics[width=0.32\textwidth]{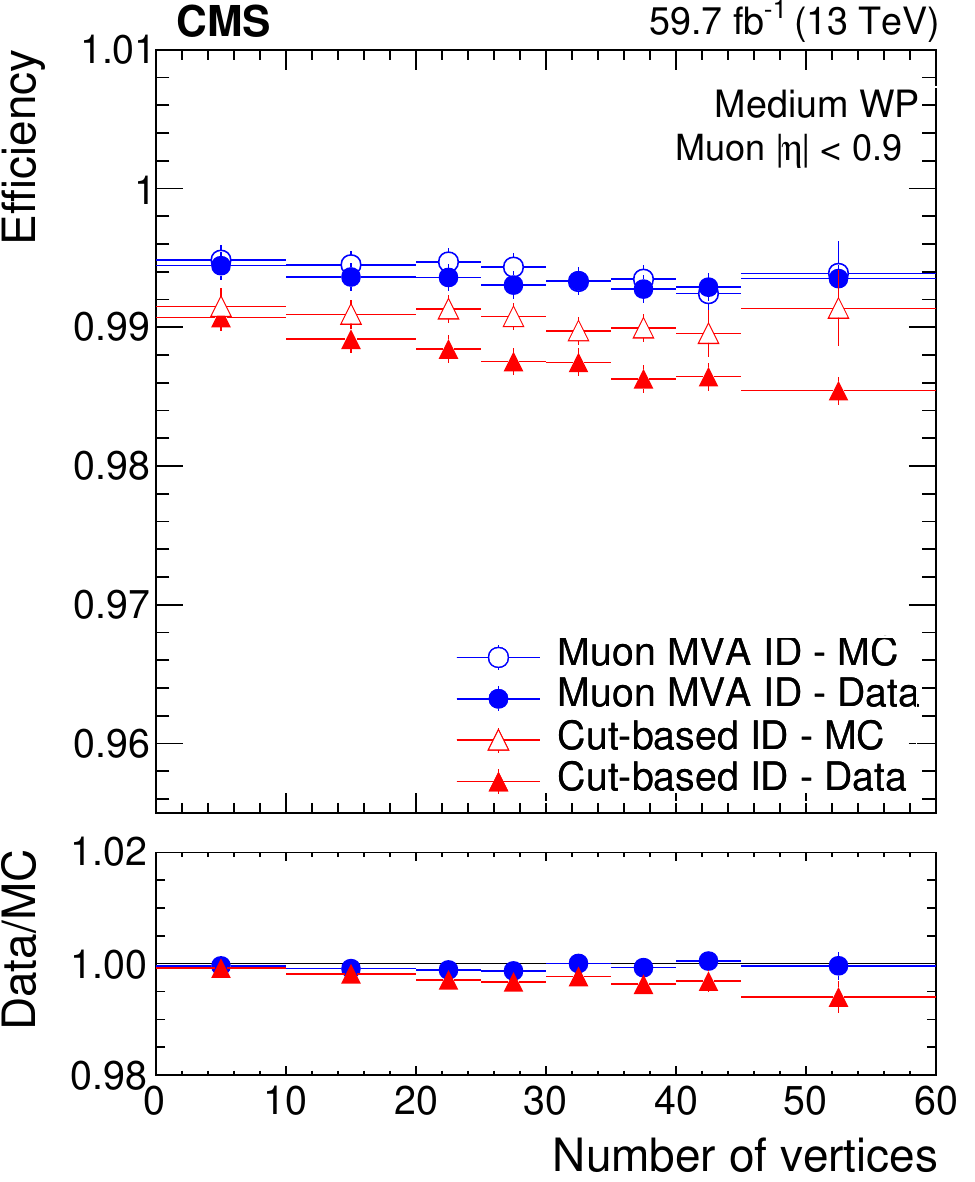}
  \includegraphics[width=0.32\textwidth]{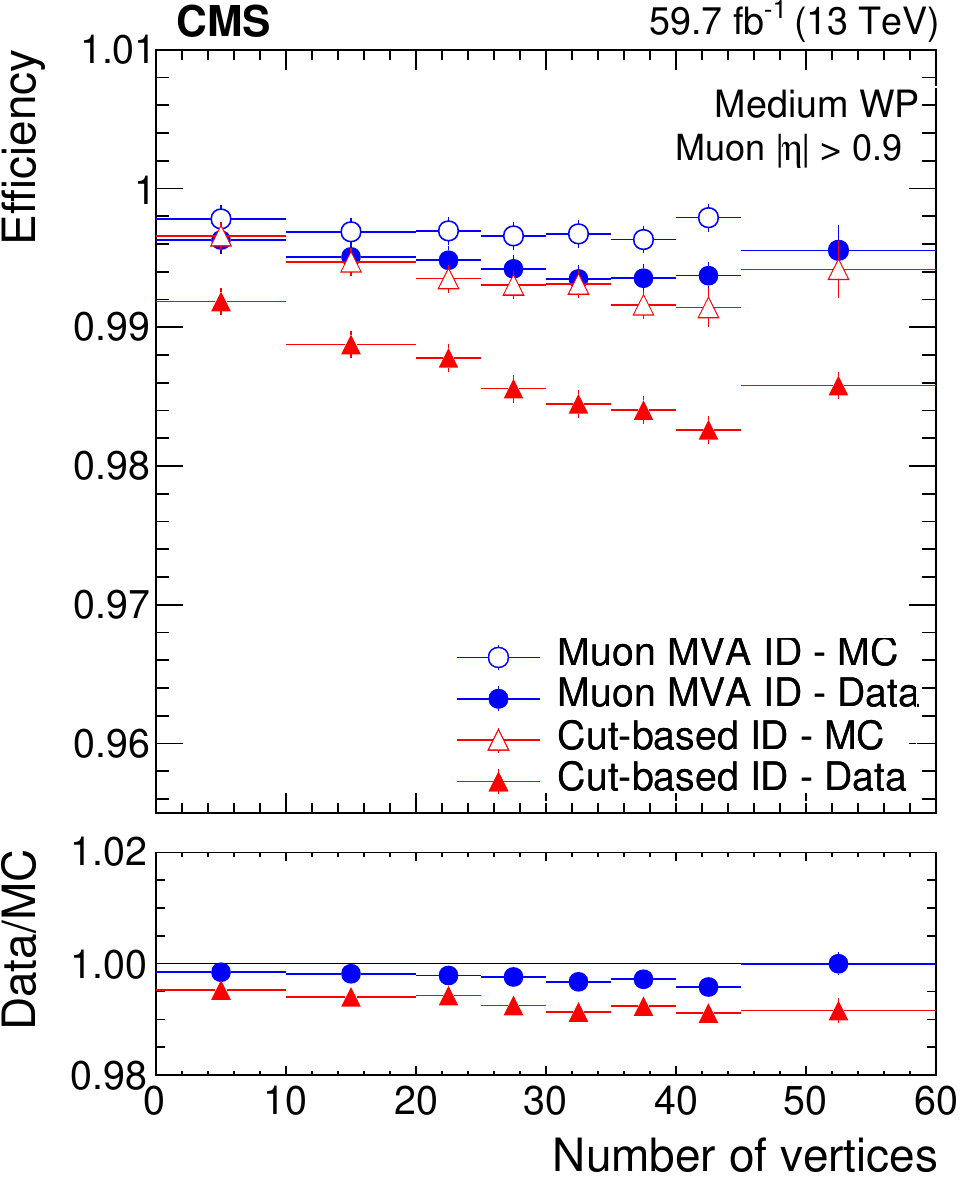} \\
  \includegraphics[width=0.32\textwidth]{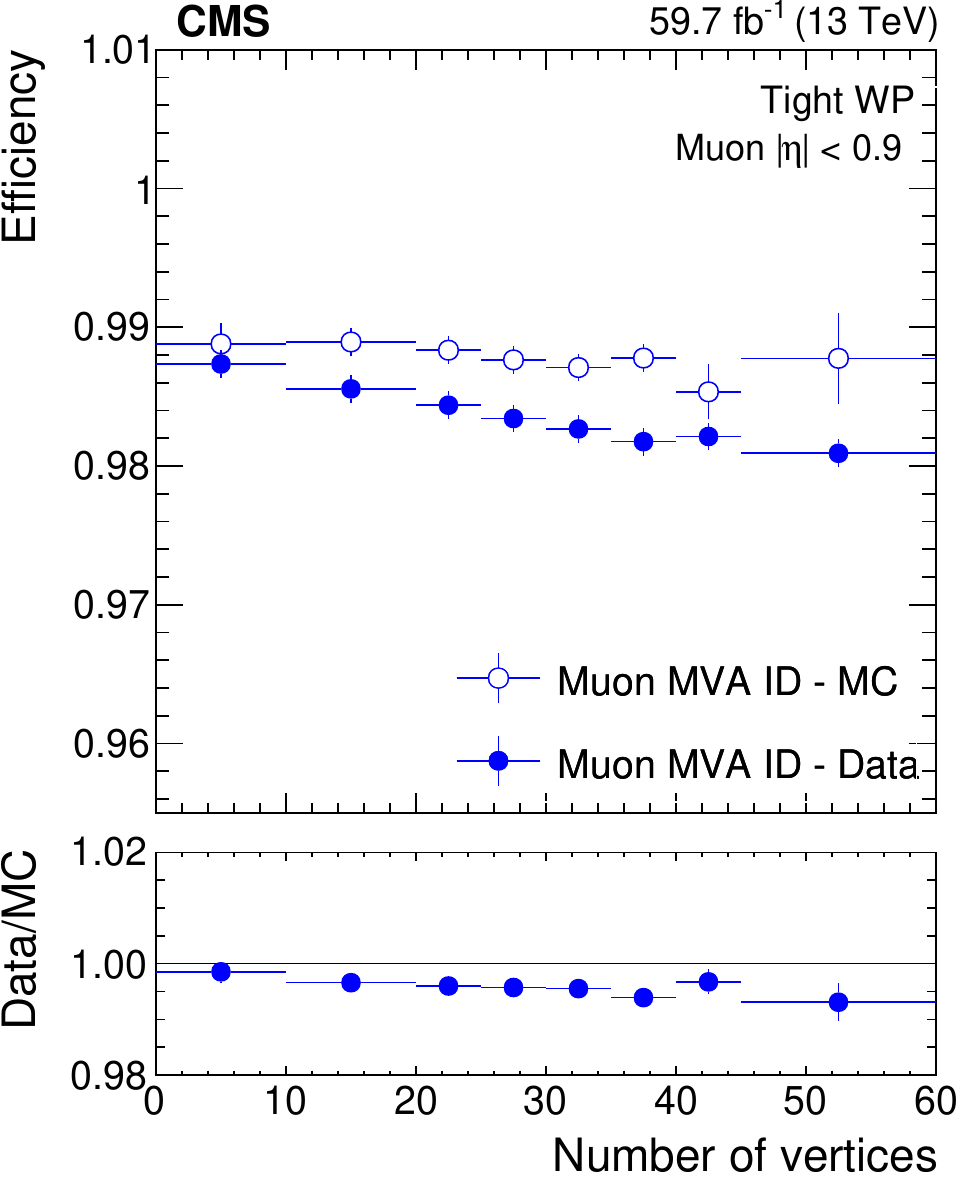}
  \includegraphics[width=0.32\textwidth]{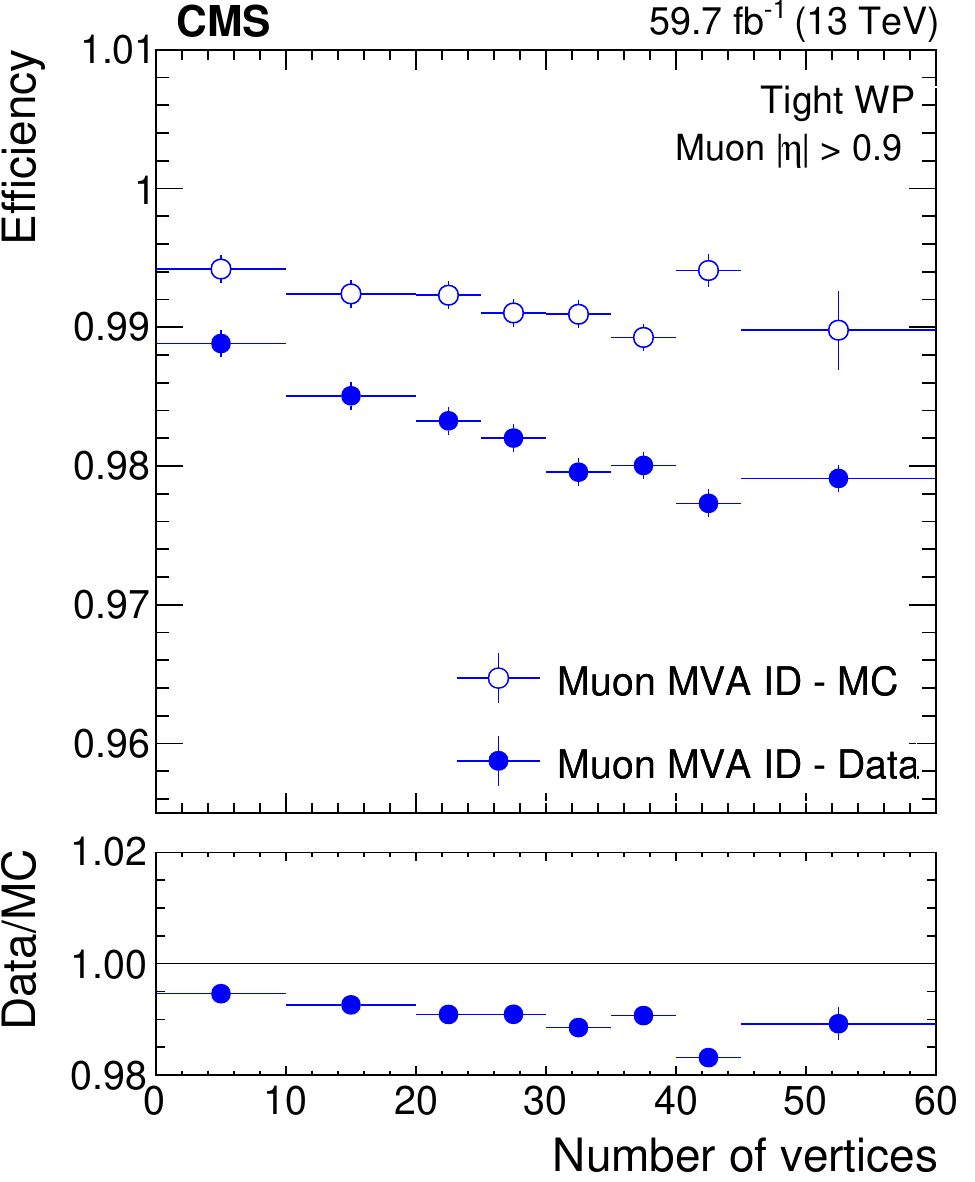}
  \caption{Muon identification efficiency for the medium (upper) and tight (lower) WPs as a function of PU for muons with $\abseta < 0.9$ (left) and $\abseta > 0.9$  (right). Blue dots show the muon MVA ID performance both for the 2018 data set and DY simulation, whereas red triangles show the efficiency of the medium cut-based ID used during \Run2. The data to MC ratio is also shown.}
  \label{fig:EffvsPU}
\end{figure}

The measured efficiencies as functions of \pt for the medium and tight WPs are shown in Fig.~\ref{fig:EffvsPt} in two $\eta$ regions. The efficiency of the medium MVA ID is higher than 99.5\%, and about 0.5\% higher than the one achieved by the cut-based ID for a similar background contamination of around 50\%. For the tight MVA ID WP, it achieves a 10\% smaller background contamination than the medium MVA ID, and the efficiency is about 99\% for muons with $\pt > 30\GeV$ and 1--2\% smaller for muons with \pt between 10 and 30\GeV. Efficiencies in the \pt region between 120 and 200\GeV are evaluated using DY simulation and they are compatible with the ones in the 80 to 120\GeV region, but they show an increase in the uncertainty of approximately a factor 2. Figure~\ref{fig:EffvsPU} shows the efficiency as a function of the number of vertices for both WPs. It remains around 99.5 (98.0)\% for the medium (tight) muon MVA ID WP even for events with up to 60 PU vertices. For the medium WP of the cut-based ID, the efficiency decreases as a function of the number of PU vertices, whereas it is stable for the medium WP of the muon MVA ID.
The efficiency in DY simulation is systematically 0.5--1.0\% higher than the efficiency in data as a result of small imperfections in the modelling. The discrepancy is observed both in the cut-based ID, as reported in Ref.~\cite{CMS:2018rym}, and in the muon MVA ID.

\subsection{Prompt-muon MVA performance}

The efficiency of the selected WP (prompt-muon MVA${}> 0.85$) is measured both in data and DY simulation as a function of the muon \pt in two $\eta$ regions using the tag-and-probe method, as in the previous section.

\begin{figure}[!htp]
  \centering
  \includegraphics[width=0.32\textwidth]{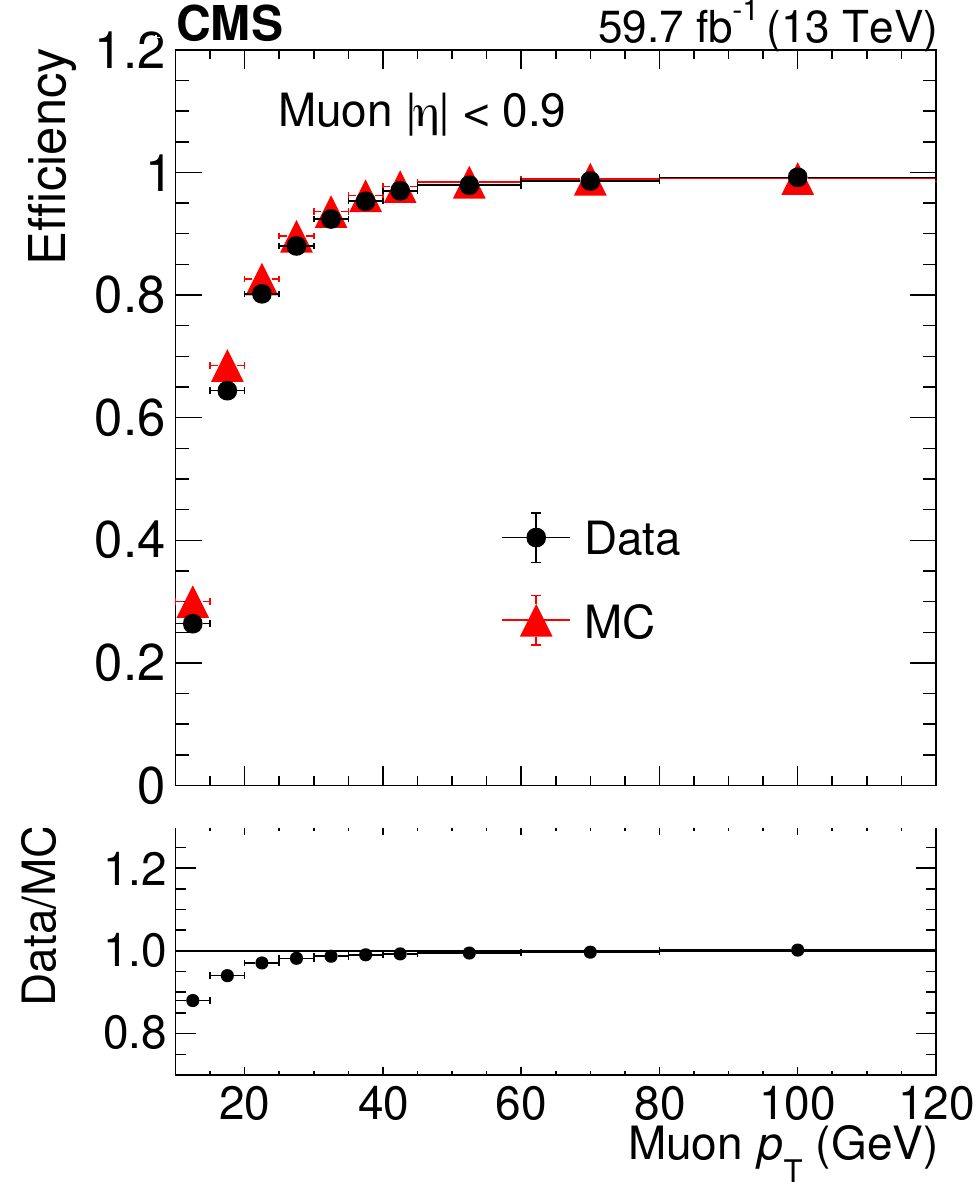}
  \includegraphics[width=0.32\textwidth]{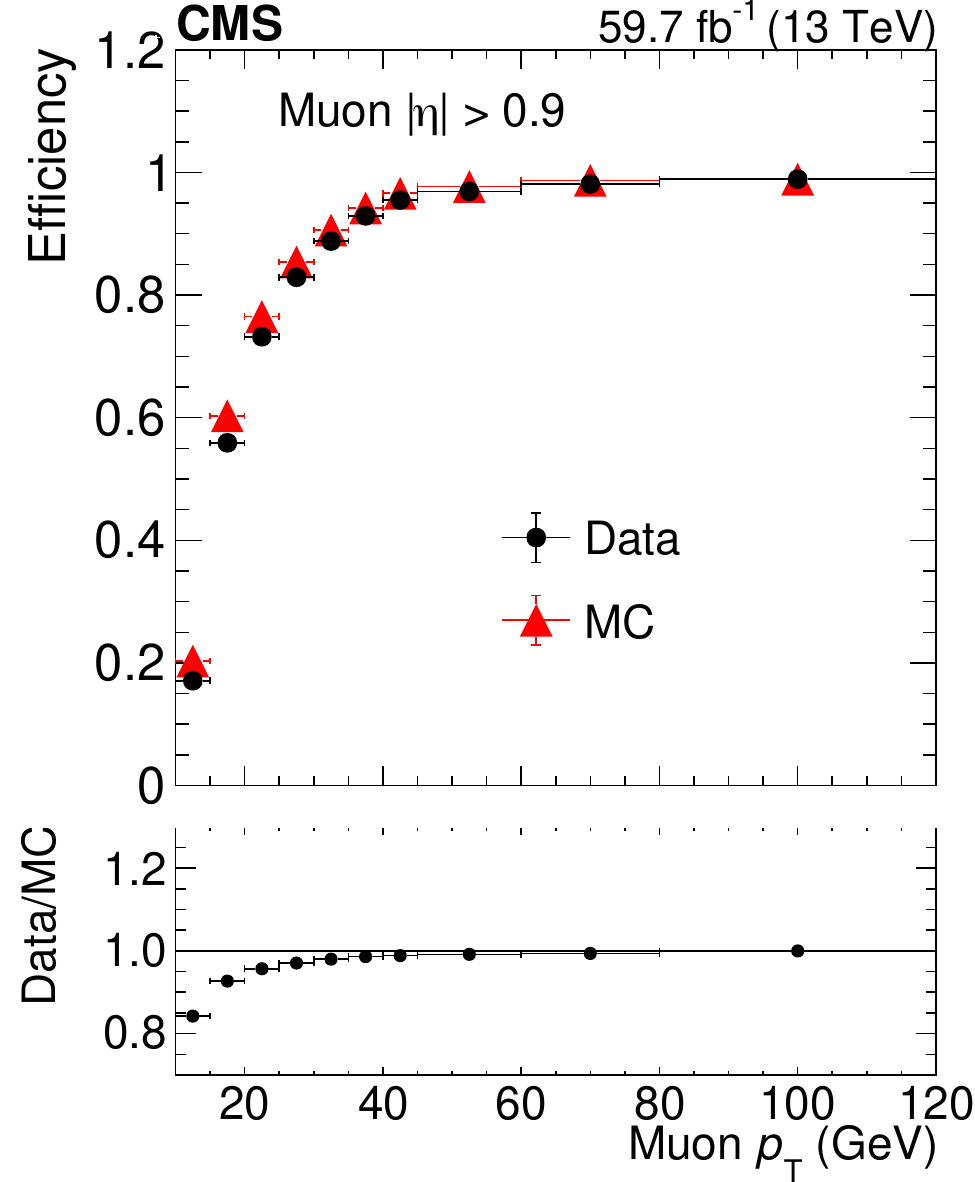}
  \caption{Efficiency of the prompt-muon MVA selection as a function of the muon \pt for muons with $\abseta < 0.9$ (left) and  $\abseta > 0.9$  (right), for the 2018 data set with black dots and simulated DY events with red triangles. The vertical bars on the plots represent the statistical uncertainty of each measurement.}
  \label{fig:lepMVA_tnp_efficiency}
\end{figure}

The measured efficiencies are shown in Fig.~\ref{fig:lepMVA_tnp_efficiency}. Efficiencies
are above 80\% for  muons with $\pt > 20\GeV$. The efficiency is modelled well by simulations, with discrepancies
smaller than 3\% for muons with $\pt > 20\GeV$. Efficiencies in the \pt region between 120 and 200\GeV are evaluated using \ttbar simulation and are compatible with the values in the 80 to 120\GeV region.

\begin{figure}[!ht]
  \centering
  \includegraphics[width=0.3\textwidth]{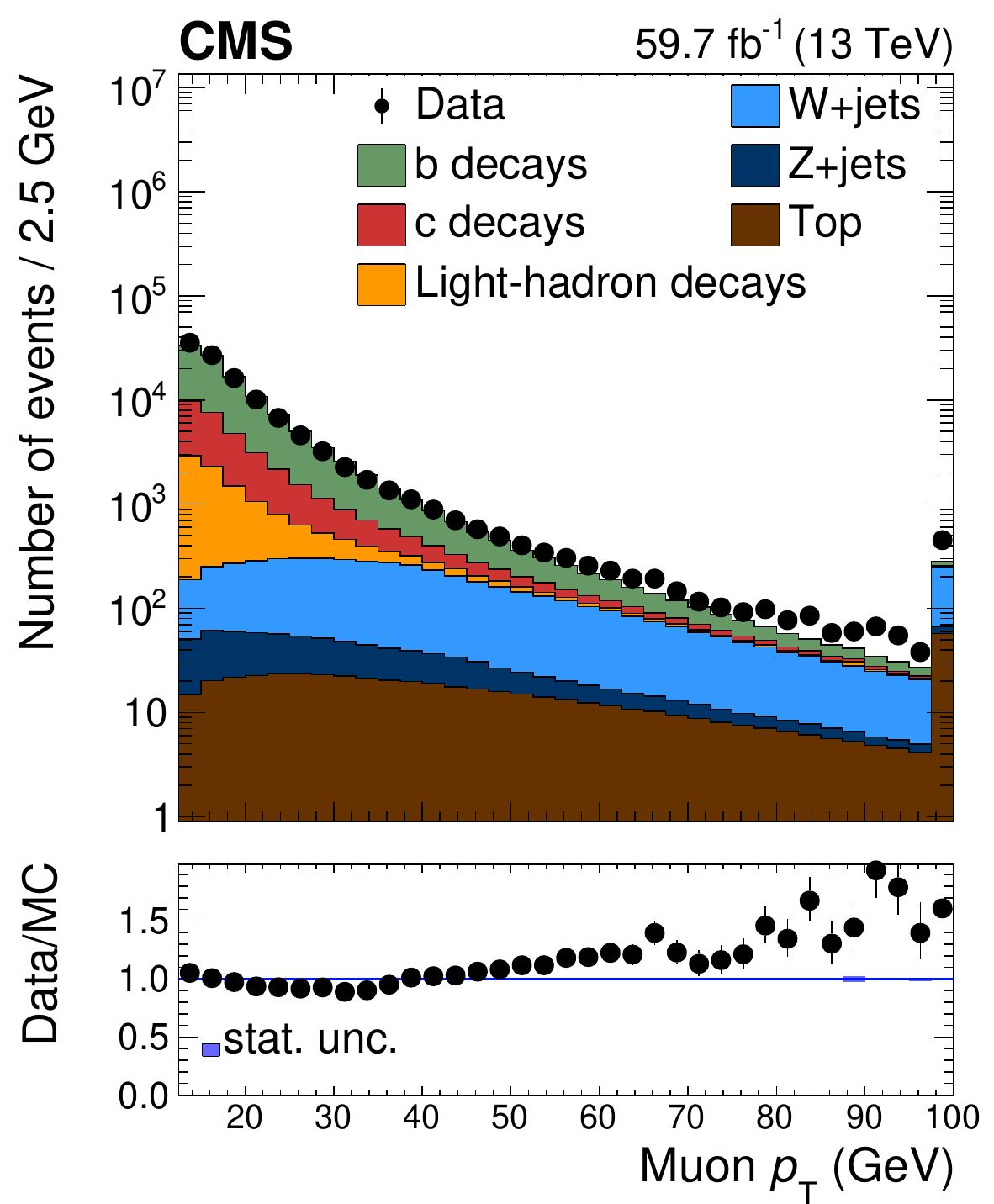}
  \includegraphics[width=0.3\textwidth]{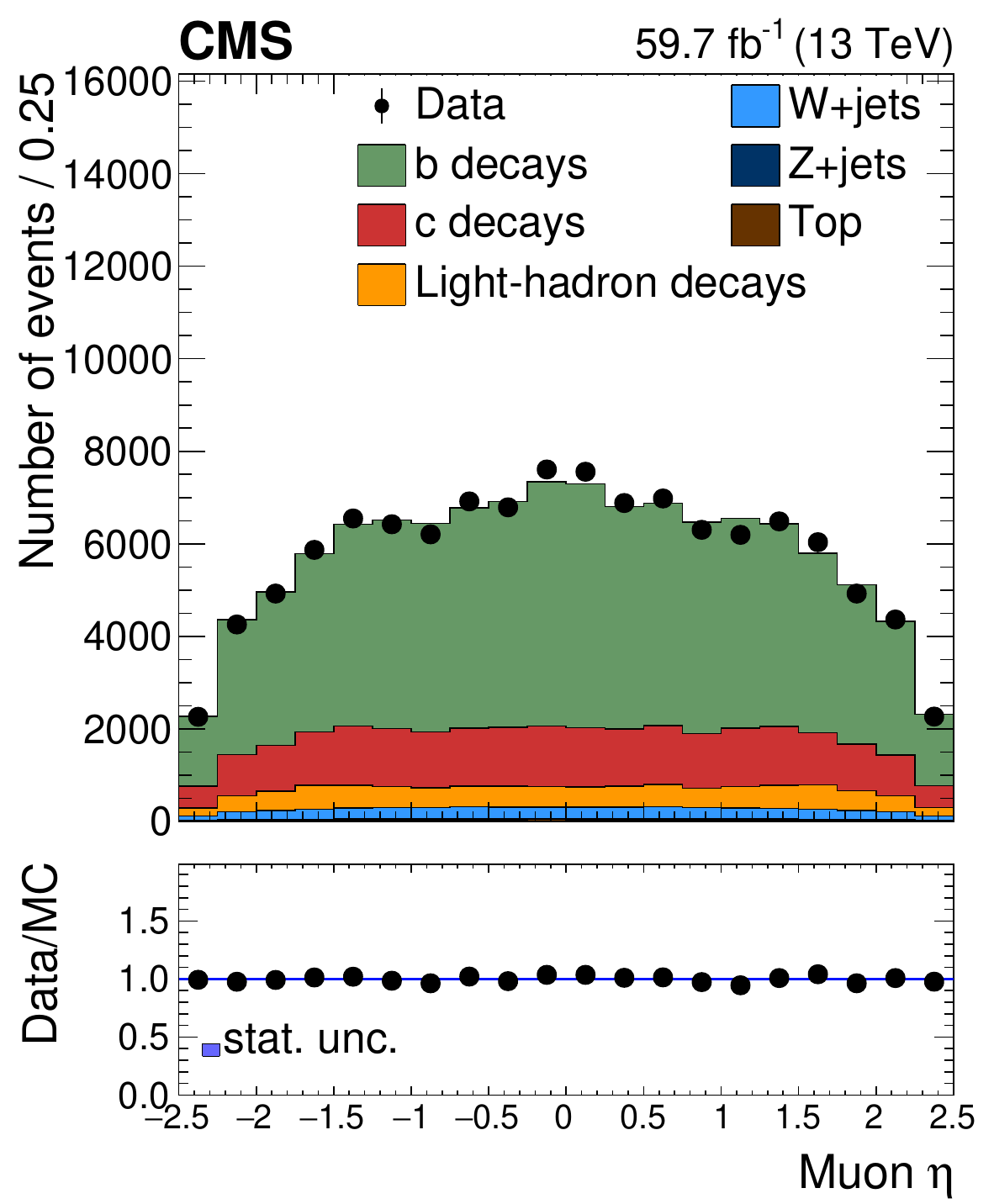}
  \includegraphics[width=0.3\textwidth]{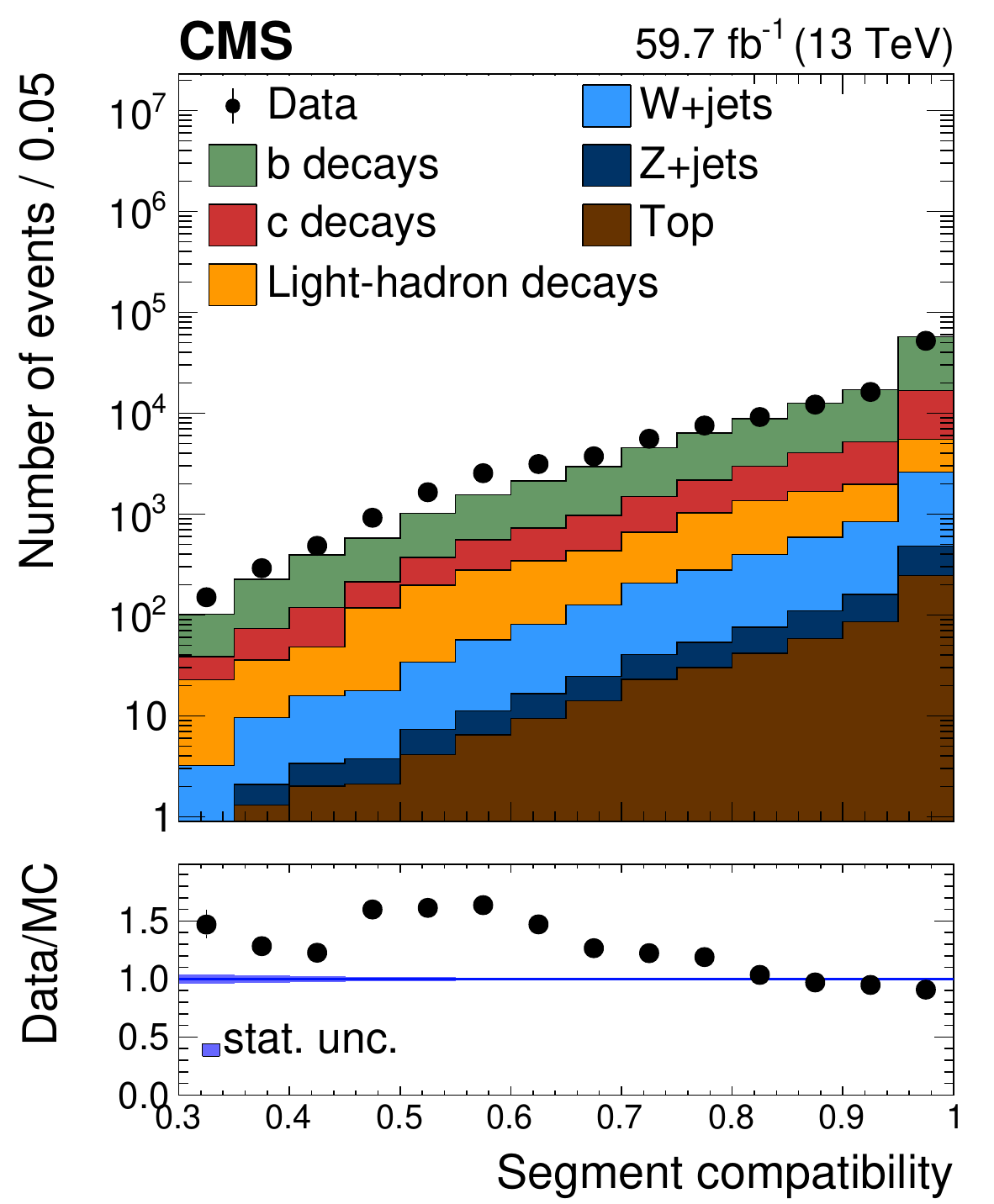} \\
  \includegraphics[width=0.3\textwidth]{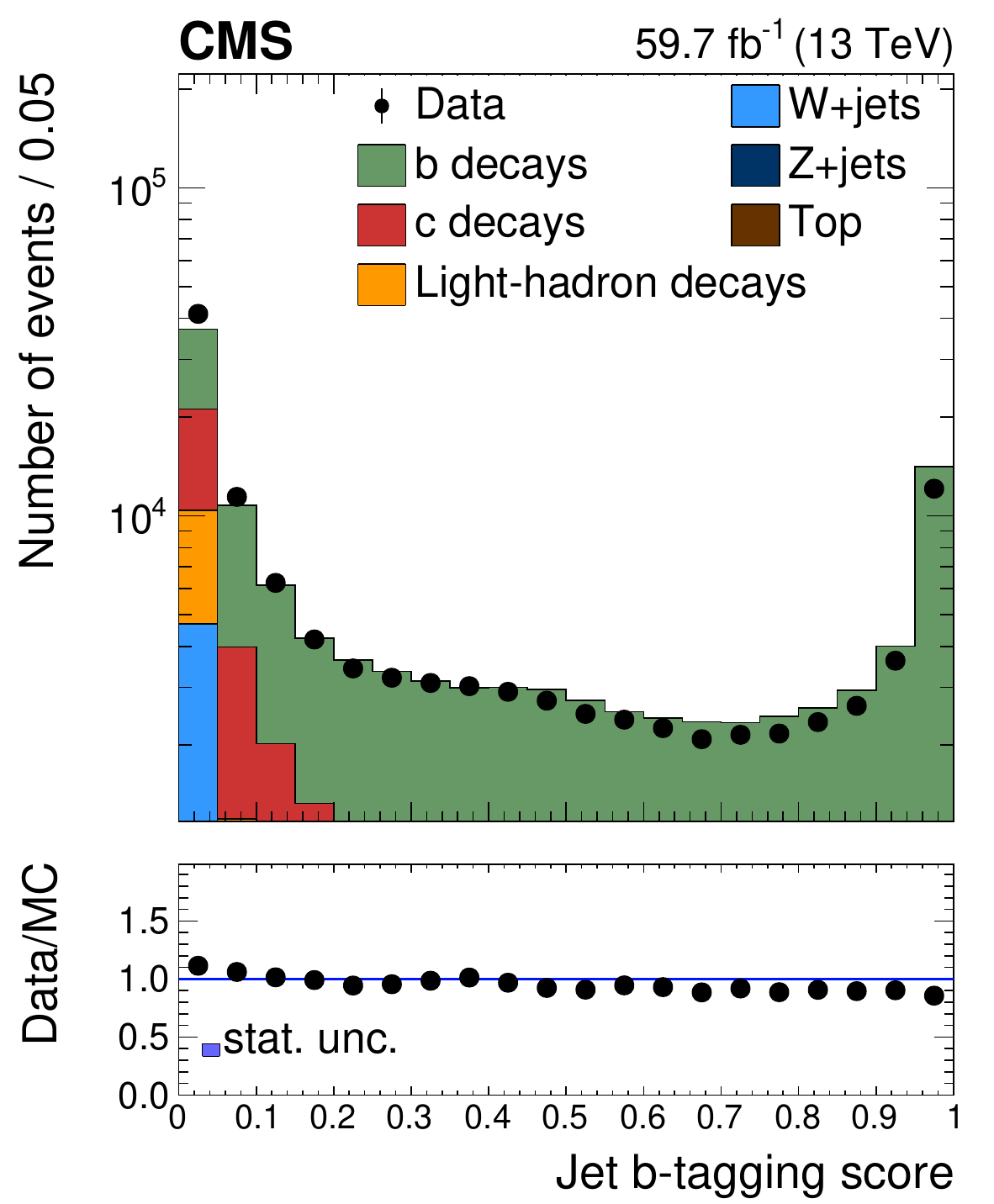}
  \includegraphics[width=0.3\textwidth]{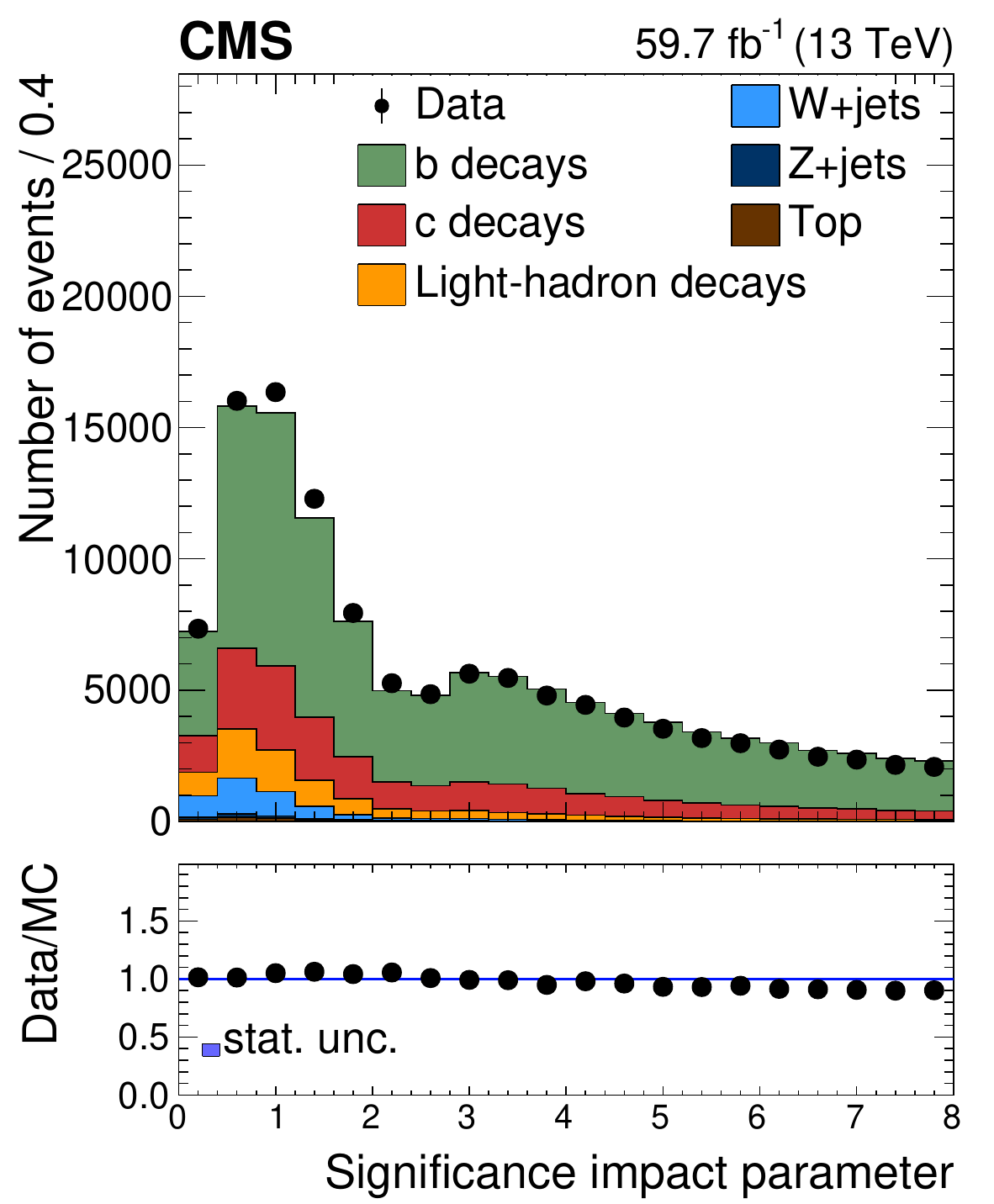}
  \includegraphics[width=0.3\textwidth]{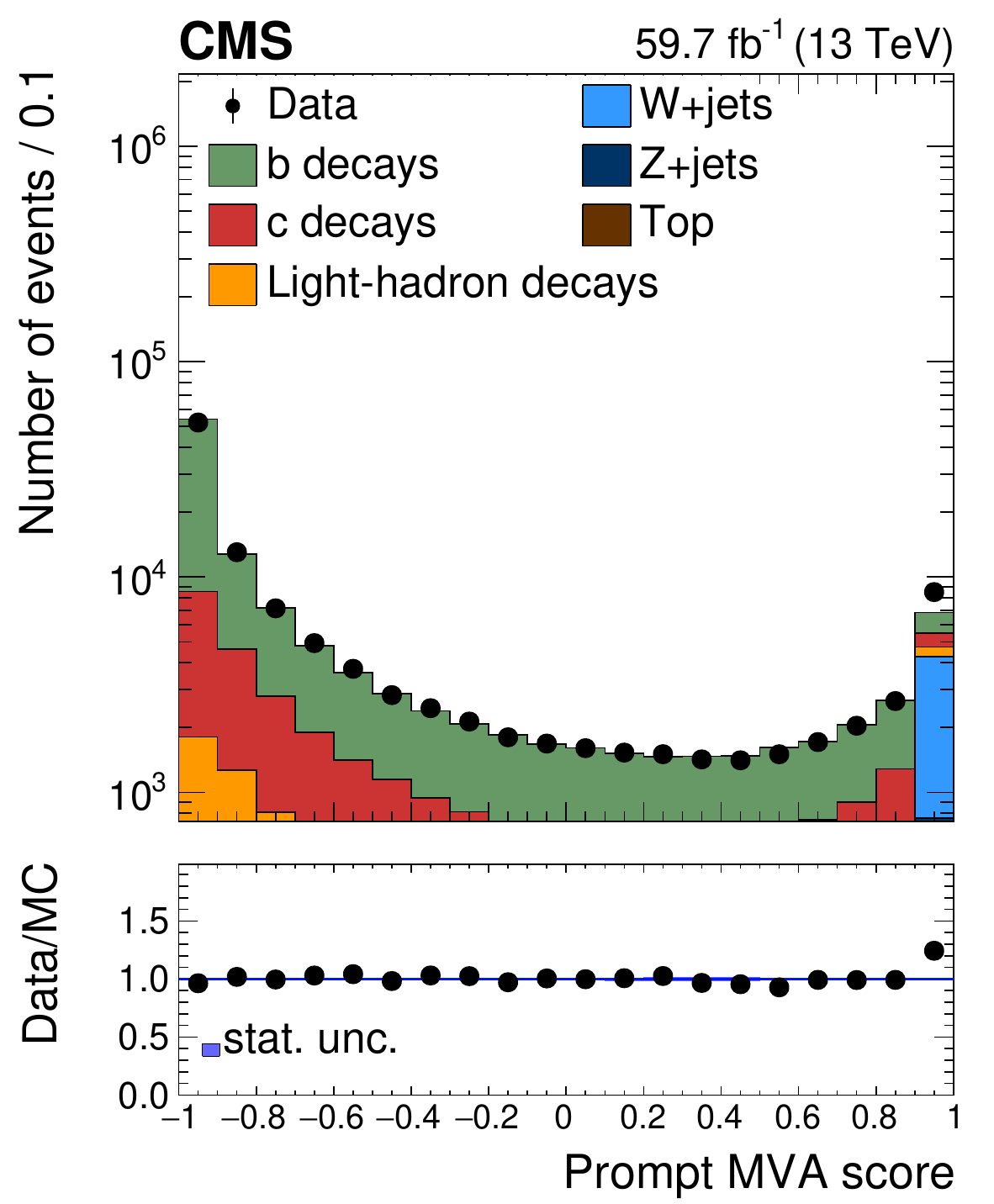}
 \caption{Distribution of data in the multijet control region as a function of the muon \pt (upper left), $\eta$ (upper central), segment compatibility (upper right), the \DeepJet \PQb tagging score of the jet associated to the muon (lower left), the significance of the impact parameter (lower central) and the prompt-muon MVA score (lower right). The vertical bars on the dots represent the statistical uncertainty of each data point and the blue band, the uncertainty associated to the limited number of simulated events.}
  \label{fig:leptonMVA_controlplots}
\end{figure}

The rate of nonprompt muons is measured in a sample enriched in multijet events. This sample is obtained selecting data collected with a set of prescaled single muon triggers that do not require isolation. The same trigger selection is applied to the simulated events. Events are required to have a muon
passing the minimal muon selection described in Section~\ref{sec:mva}. Additionally, to suppress
the contribution from light-flavour meson decays, muons are required to fulfill the medium cut-based requirements.
Moreover, events are required to have a jet with $\pt > 30\GeV$ and $\abseta < 2.4$ recoiling against a muon with a $\DR$ between the jet and the muon of less than 0.7. Figure~\ref{fig:leptonMVA_controlplots} shows a set of the most important variables for events passing this selection. The multijet background enriched region defined above is dominated by nonprompt muons mostly coming from heavy-flavour hadron decays. In general,
the input variables of the prompt-muon MVA, as well as the MVA score, are modelled quite well by simulation.
An exception to this is the segment compatibility variable, which shows a clear disagreement
at lower values. This is due to the fact that simulated multijet events are filtered, so they contain a
muon produced at the PV, and they do not include a significant part of the PU contribution because only events with a muon in the hard scattering are selected. These muons are typically not relevant for the analyses, since they are easily rejected by the preselection requirements. The \pt distribution is also not well modelled in these background samples,
because the employed multijet samples have a limited accuracy. However, the amount of mismodelling is small
in the signal and background samples used in the training.

The sample defined by the selection described above is enriched in nonprompt muons. However, background contributions may arise from electroweak-induced processes such as \Wjets or \ttbar events. To subtract this component, the observable \mtfix is defined as:
\begin{linenomath}\begin{equation}
\mtfix = \sqrt{2 \ptfix \, \ptmiss \, ( 1 - \cos \Delta\phi )},
\label{eq:mTfix}
\end{equation}\end{linenomath}
where $\Delta\phi$ is the azhimutal angle between the muon and \ptvecmiss.
The definition of \mtfix is similar to that of the transverse mass of the muon and
\ptmiss; the only difference is that the muon momentum is set to $\ptfix = 35\GeV$, approximately the
mode of the muon \pt distribution in \Wjets events. Similarly to the
transverse mass, the \mtfix variable has a
kinematic endpoint near the \PW boson mass. However, since the muon \pt is not included
in this calculation, it is less correlated with this quantity, avoiding potential biases in the
measurement.

Then, the number of events with a nonprompt muon is computed by means of a fit to the \mtfix observable. In the fit, the prompt and nonprompt contributions are parameterized using templates that are derived
from samples of simulated events of each muon source. The prompt contribution is obtained from \Wjets, DY$+$jets,
and \ttbar simulated events, whereas the nonprompt contribution includes multijet events with muons in the final state.
The fit also incorporates nuisance
parameters to account for statistical uncertainties in the templates, as well as systematic
deformations of their shapes. The fit is performed separately for cases in which the muon
passes or fails a prompt-muon MVA score larger than 0.85. We label the result of each of
the two fits as \Npass and \Nfail, respectively, and the nonprompt rate, $f$, is computed as
\begin{linenomath}\begin{equation}
  f = \frac{\Npass}{\Npass + \Nfail}.
\end{equation}\end{linenomath}
Additionally, to compare the performance of the prompt-muon MVA with a more standard criterion designed
to reject \ttbar events, the nonprompt-muon rate with a selection based on mini-isolation is also computed.
Since the prompt-muon MVA and the mini-isolation have a different efficiency dependence as a function of
the muon \pt, a requirement on mini-isolation, which is dependent on \pt, is included that gives the same efficiency as the prompt-muon MVA and thus provides a fair comparison.

\begin{figure}[!b]
  \centering
  \includegraphics[width=0.3\textwidth]{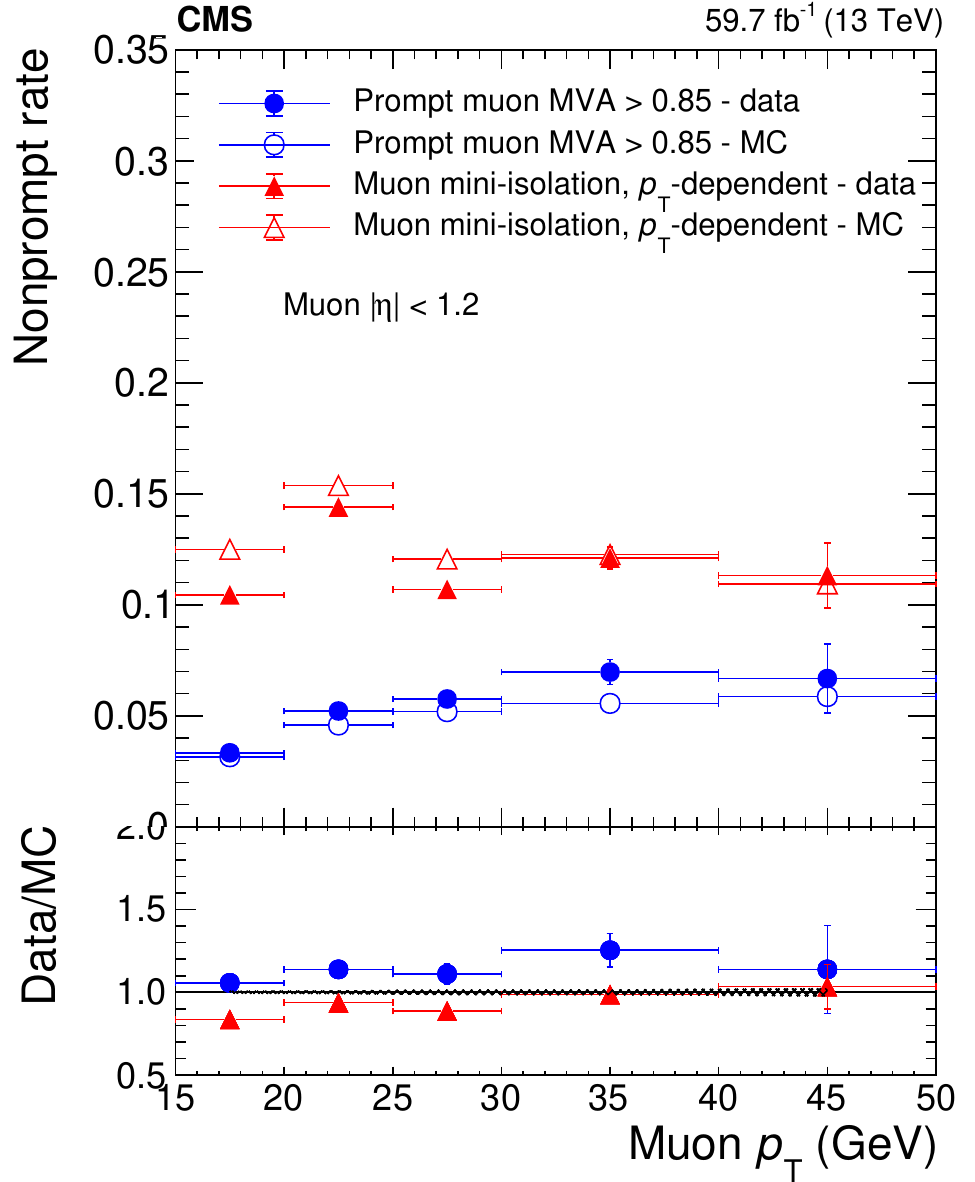}
  \includegraphics[width=0.3\textwidth]{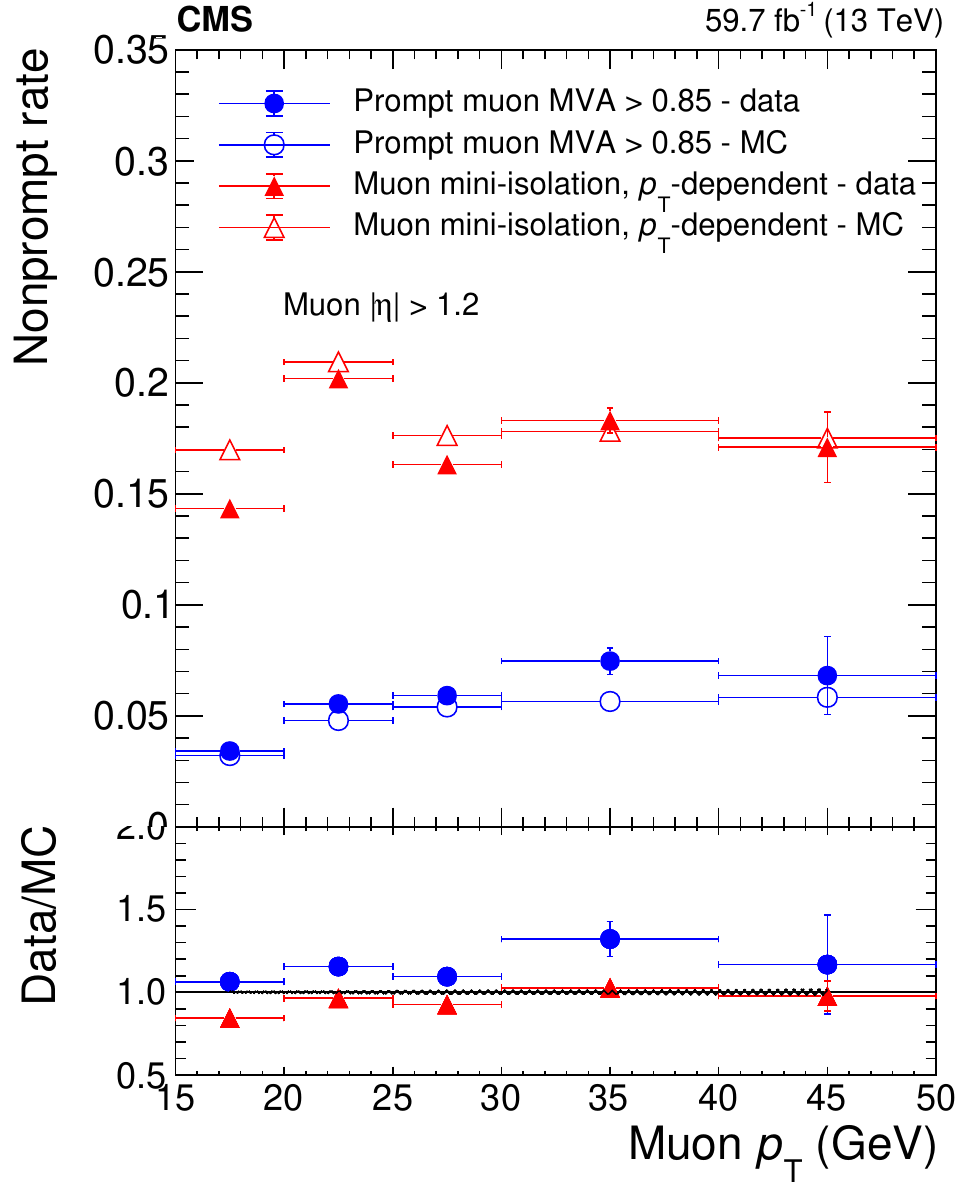}
  \caption{Measurement of the nonprompt-muon rate of a prompt-muon MVA (blue dots) and mini-isolation (red triangles) selection
    as a function of \pt for muons with $\abseta < 1.2$ (left) and $\abseta > 1.2$ (right) for the 2018 data set and simulated DY events. }
  \label{fig:muonFRMultijet}
\end{figure}

The nonprompt-muon rates for the two selections are shown in Fig.~\ref{fig:muonFRMultijet}, displaying approximately a factor 2 (3) smaller
nonprompt rate of the prompt-muon MVA with respect to the mini-isolation selection for muons with $\abseta < 1.2$ ($>$1.2).
Additionally, the observed nonprompt rate agrees with the predictions provided by the
simulations, which also reproduces the difference in nonprompt rate between the two strategies. This result
validates the prompt-muon MVA in collision data and also shows its generalization power in a phase space different
from the one used for its training.

\section{Summary}
\label{sec:summary}

A correct identification of the leptons is crucial in many precision measurements and searches to suppress the otherwise overwhelming background and as an indicator of interesting physical processes. Two multivariate analyses were developed for highly efficient muon identification and isolation. The first one, the muon MVA ID, is trained to distinguish muons produced promptly in heavy gauge boson decays as well as muons from \PGt lepton and heavy-flavour hadron decays, from background muons produced in light-hadron decays (pions or kaons) or other spurious signatures in the detector that could be misreconstructed as muons. The discriminator is presented as an alternative to the standard cut-based identification criteria and could be used for high-efficiency working points. The second one, the prompt-muon MVA, selects isolated muons from \PW, \PZ, Higgs bosons, and \PGt lepton decays to reduce the contamination from nonisolated muons arising in heavy-flavour hadron decays. Their performances are measured in proton-proton collisions recorded by the CMS experiment during 2018 and compared to simulation. The performance of the muon MVA ID improves significantly that of the cut-based ID and the prompt-muon MVA achieves a factor 2--3 times smaller nonprompt-muon rates than the mini-isolation selection.

\begin{acknowledgments}
We congratulate our colleagues in the CERN accelerator departments for the excellent performance of the LHC and thank the technical and administrative staffs at CERN and at other CMS institutes for their contributions to the success of the CMS effort. In addition, we gratefully acknowledge the computing centres and personnel of the Worldwide LHC Computing Grid and other centres for delivering so effectively the computing infrastructure essential to our analyses. Finally, we acknowledge the enduring support for the construction and operation of the LHC, the CMS detector, and the supporting computing infrastructure provided by the following funding agencies: SC (Armenia), BMBWF and FWF (Austria); FNRS and FWO (Belgium); CNPq, CAPES, FAPERJ, FAPERGS, and FAPESP (Brazil); MES and BNSF (Bulgaria); CERN; CAS, MoST, and NSFC (China); MINCIENCIAS (Colombia); MSES and CSF (Croatia); RIF (Cyprus); SENESCYT (Ecuador); MoER, ERC PUT and ERDF (Estonia); Academy of Finland, MEC, and HIP (Finland); CEA and CNRS/IN2P3 (France); SRNSF (Georgia); BMBF, DFG, and HGF (Germany); GSRI (Greece); NKFIH (Hungary); DAE and DST (India); IPM (Iran); SFI (Ireland); INFN (Italy); MSIP and NRF (Republic of Korea); MES (Latvia); LAS (Lithuania); MOE and UM (Malaysia); BUAP, CINVESTAV, CONACYT, LNS, SEP, and UASLP-FAI (Mexico); MOS (Montenegro); MBIE (New Zealand); PAEC (Pakistan); MES and NSC (Poland); FCT (Portugal);  MESTD (Serbia); MCIN/AEI and PCTI (Spain); MOSTR (Sri Lanka); Swiss Funding Agencies (Switzerland); MST (Taipei); MHESI and NSTDA (Thailand); TUBITAK and TENMAK (Turkey); NASU (Ukraine); STFC (United Kingdom); DOE and NSF (USA).

\hyphenation{Rachada-pisek} Individuals have received support from the Marie-Curie programme and the European Research Council and Horizon 2020 Grant, contract Nos.\ 675440, 724704, 752730, 758316, 765710, 824093, and COST Action CA16108 (European Union); the Leventis Foundation; the Alfred P.\ Sloan Foundation; the Alexander von Humboldt Foundation; the Science Committee, project no. 22rl-037 (Armenia); the Belgian Federal Science Policy Office; the Fonds pour la Formation \`a la Recherche dans l'Industrie et dans l'Agriculture (FRIA-Belgium); the Agentschap voor Innovatie door Wetenschap en Technologie (IWT-Belgium); the F.R.S.-FNRS and FWO (Belgium) under the ``Excellence of Science -- EOS" -- be.h project n.\ 30820817; the Beijing Municipal Science \& Technology Commission, No. Z191100007219010 and Fundamental Research Funds for the Central Universities (China); the Ministry of Education, Youth and Sports (MEYS) of the Czech Republic; the Shota Rustaveli National Science Foundation, grant FR-22-985 (Georgia); the Deutsche Forschungsgemeinschaft (DFG), under Germany's Excellence Strategy -- EXC 2121 ``Quantum Universe" -- 390833306, and under project number 400140256 - GRK2497; the Hellenic Foundation for Research and Innovation (HFRI), Project Number 2288 (Greece); the Hungarian Academy of Sciences, the New National Excellence Program - \'UNKP, the NKFIH research grants K 124845, K 124850, K 128713, K 128786, K 129058, K 131991, K 133046, K 138136, K 143460, K 143477, 2020-2.2.1-ED-2021-00181, and TKP2021-NKTA-64 (Hungary); the Council of Science and Industrial Research, India; ICSC -- National Research Centre for High Performance Computing, Big Data and Quantum Computing, funded by the EU NexGeneration program (Italy); the Latvian Council of Science; the Ministry of Education and Science, project no. 2022/WK/14, and the National Science Center, contracts Opus 2021/41/B/ST2/01369 and 2021/43/B/ST2/01552 (Poland); the Funda\c{c}\~ao para a Ci\^encia e a Tecnologia, grant CEECIND/01334/2018 (Portugal); the National Priorities Research Program by Qatar National Research Fund; MCIN/AEI/10.13039/501100011033, ERDF ``a way of making Europe", and the Programa Estatal de Fomento de la Investigaci{\'o}n Cient{\'i}fica y T{\'e}cnica de Excelencia Mar\'{\i}a de Maeztu, grant MDM-2017-0765 and Programa Severo Ochoa del Principado de Asturias (Spain); the Chulalongkorn Academic into Its 2nd Century Project Advancement Project, and the National Science, Research and Innovation Fund via the Program Management Unit for Human Resources \& Institutional Development, Research and Innovation, grant B05F650021 (Thailand); the Kavli Foundation; the Nvidia Corporation; the SuperMicro Corporation; the Welch Foundation, contract C-1845; and the Weston Havens Foundation (USA).
\end{acknowledgments}

\bibliography{auto_generated}
\cleardoublepage \appendix\section{The CMS Collaboration \label{app:collab}}\begin{sloppypar}\hyphenpenalty=5000\widowpenalty=500\clubpenalty=5000
\cmsinstitute{Yerevan Physics Institute, Yerevan, Armenia}
{\tolerance=6000
A.~Hayrapetyan, A.~Tumasyan\cmsAuthorMark{1}\cmsorcid{0009-0000-0684-6742}
\par}
\cmsinstitute{Institut f\"{u}r Hochenergiephysik, Vienna, Austria}
{\tolerance=6000
W.~Adam\cmsorcid{0000-0001-9099-4341}, J.W.~Andrejkovic, T.~Bergauer\cmsorcid{0000-0002-5786-0293}, S.~Chatterjee\cmsorcid{0000-0003-2660-0349}, K.~Damanakis\cmsorcid{0000-0001-5389-2872}, M.~Dragicevic\cmsorcid{0000-0003-1967-6783}, A.~Escalante~Del~Valle\cmsorcid{0000-0002-9702-6359}, P.S.~Hussain\cmsorcid{0000-0002-4825-5278}, M.~Jeitler\cmsAuthorMark{2}\cmsorcid{0000-0002-5141-9560}, N.~Krammer\cmsorcid{0000-0002-0548-0985}, D.~Liko\cmsorcid{0000-0002-3380-473X}, I.~Mikulec\cmsorcid{0000-0003-0385-2746}, J.~Schieck\cmsAuthorMark{2}\cmsorcid{0000-0002-1058-8093}, R.~Sch\"{o}fbeck\cmsorcid{0000-0002-2332-8784}, D.~Schwarz\cmsorcid{0000-0002-3821-7331}, M.~Sonawane\cmsorcid{0000-0003-0510-7010}, S.~Templ\cmsorcid{0000-0003-3137-5692}, W.~Waltenberger\cmsorcid{0000-0002-6215-7228}, C.-E.~Wulz\cmsAuthorMark{2}\cmsorcid{0000-0001-9226-5812}
% ...
\par}
\vskip\cmsinstskip
\dag:~Deceased\\
$^{1}$Also at Yerevan State University, Yerevan, Armenia\\
$^{2}$Also at TU Wien, Vienna, Austria\\
% ...
\end{sloppypar}
\end{document}